\documentclass[conference]{IEEEtran}
\IEEEoverridecommandlockouts

\usepackage{cite}
\usepackage{amsmath,amssymb,amsfonts}
\usepackage{algorithm}
\usepackage{algorithmic}
\usepackage{graphicx}
\usepackage{textcomp}
\usepackage[table]{xcolor}
\usepackage{booktabs}
\usepackage{multirow}
\usepackage{pifont}
\usepackage{enumitem}
\usepackage{url}
\definecolor{refblue}{RGB}{0,0,255}
\usepackage[implicit=false]{hyperref}
\hypersetup{hidelinks}          
\newcommand{\codeurl}{\textcolor{blue}{\url{https://github.com/SeoLabCornell/MiX}}}
\usepackage{eso-pic}

\AtBeginDocument{%
  \let\mixorigref\ref
  \renewcommand{\ref}[1]{\textcolor{refblue}{\mixorigref{#1}}}%
}

\makeatletter
\def\section{\@startsection{section}{1}{\z@}{3.50ex plus 0.3ex minus 0.1ex}%
{0.9ex plus 0.2ex minus 0ex}{\normalfont\normalsize\centering\scshape}}%
\def\subsection{\@startsection{subsection}{2}{\z@}{2.80ex plus 0.3ex minus 0.1ex}%
{0.9ex plus 0.2ex minus 0ex}{\normalfont\normalsize\itshape}}%
\makeatother

\newlength{\fpad}

\newcommand{\xmark}{\ding{55}}

\begin{document}

\title{MiX: Micro-Inverted-Scaling for End-to-End\\
Low-Bit Vision-Language Model Acceleration}

\author{\IEEEauthorblockN{Yuan Liao}
\IEEEauthorblockA{\textit{Cornell Tech, Cornell University} \\
New York, NY, USA \\
yl3662@cornell.edu}
\and
\IEEEauthorblockN{Jae-sun Seo}
\IEEEauthorblockA{\textit{Cornell Tech, Cornell University} \\
New York, NY, USA \\
js3528@cornell.edu}
}

\AddToShipoutPictureBG*{%
  \AtPageLowerLeft{%
    \put(0,\LenToUnit{0.45in}){\makebox[\paperwidth][c]{\footnotesize
      This paper has been accepted to the 59th IEEE/ACM International Symposium on Microarchitecture (MICRO 2026).}}}}
\maketitle

\begin{abstract}
The deployment of Vision-Language Models (VLMs) on edge devices is severely bottlenecked by memory bandwidth, necessitating aggressive sub-8-bit quantization. Since edge accelerators are strictly constrained by area and power, they require end-to-end quantized models. However, the extreme dynamic range gap between multi-modal tokens causes standard block formats to suffer ``microscaling collapse,'' where a single massive outlier hijacks the shared exponent, underflowing surrounding elements and destroying attention maps. To break this bottleneck, we propose Micro-Inverted-Scaling (MiX), a novel format that mathematically inverts the microscaling paradigm: rather than grouping multiple mantissas under one shared exponent, MiX groups private, per-element exponents under a single shared mantissa. To handle asymmetric VLM outlier topologies, we introduce an adaptive dual-format (MiX-MX) inference framework. By algebraically factoring out the shared MiX mantissa, this framework maps to a custom accelerator, replacing multipliers with efficient shifters. Evaluated end-to-end on multiple VLMs, our 4.5-bit MiX formulation exhibits equivalent or superior accuracy on multi-modal benchmarks compared to NVFP4. Simultaneously, the MiX accelerator delivers a 25\% improvement in area efficiency over the NVFP4 baseline and a 2.3--4.5\texttimes{} speedup with 1.4--2.9\texttimes{} energy reduction across models compared to the state-of-the-art accelerator \textit{Focus}, proving the inverted-scaling datapath is physically superior for efficient VLM deployment. Code is available at \codeurl.
\end{abstract}


\section{Introduction}

Generative AI has shifted from unimodal Large Language Models (LLMs) to multi-modal Vision-Language Models (VLMs) \cite{LLaVA, InternVL}, which augment autoregressive text generation with visual comprehension for end-to-end tasks like complex visual reasoning and document parsing \cite{MMMU}. As these models scale, processing dense visual tokens alongside semantic text tokens exacerbates the memory bandwidth and compute bottlenecks already inherent in transformer architectures \cite{attention}. During autoregressive decoding, edge memory traffic is dominated by two compounding bottlenecks: continuous fetching of massive model weights, and the explosive growth of the Key-Value (KV) cache \cite{PagedAttention, H2O} driven by tens of thousands of high-resolution visual tokens. This dual-sided memory wall mandates an end-to-end activation-weight quantization strategy.

However, current optimization paradigms exhibit a fundamental blind spot, addressing one side of the memory bottleneck while neglecting the other. Recent architectural efforts like \textit{Focus} \cite{focus} and other token-pruning accelerators \cite{SpAtten, HeatViT, AdapTiV} exploit token concentration to skip redundant visual activations, yet leave the memory wall of fetching massive, uncompressed VLM weights unresolved. Conversely, the industry's prevailing weight-only quantization (e.g., W4A16) \cite{GPTQ, OmniQuant} is fundamentally inadequate for VLMs: by leaving the massive visual KV cache \cite{KIVI, KVQuant} and dynamic activation footprints uncompressed, it fails to resolve the primary memory capacity bottleneck during generation. At the microarchitectural level, W4A16 inference further forces Processing Elements (PEs) to either employ costly mixed-precision multipliers \cite{LLMint8} or dynamically unscale weights back to FP16, leaving silicon area and dynamic power bound to 16-bit logic and nullifying any theoretical compute efficiency gains. Surviving the extreme power and area constraints of edge devices thus mandates end-to-end, fully quantized datapaths for acceleration of VLM models.

Yet, achieving robust sub-8-bit activation quantization in VLMs remains unsolved. Recent literature \cite{MBQ, Q-VLM, TLQ, VEQ} confirms that directly porting LLM-centric algorithms, e.g., AWQ \cite{AWQ}, SmoothQuant \cite{smoothquant}, SpinQuant \cite{spinquant}, and rotation-based quantization \cite{QuaRot}, induces severe accuracy degradation, stemming from a fundamental multi-modal impedance mismatch \cite{P4Q}. Visual features differ drastically from text tokens in both distribution and sensitivity: empirical profiling \cite{MQuant} shows that visual activations span a significantly broader dynamic range from $-20$ to $10$, whereas textual tokens remain tightly concentrated near zero. Applying homogeneous quantization scales across both modalities therefore either aggressively clips critical visual outliers or forces fine-grained textual features to mathematically underflow, destroying multi-modal accuracy.

Despite the proliferation of LLM accelerators \cite{BitMoD, M-ANT} and outlier-aware datapaths \cite{OliVe, ANT}, supporting VLM-specific activation dynamics under strict edge envelopes is underexplored. Standard microscaling (MX) \cite{microscaling} derivatives, building on block floating-point lineages \cite{Flexpoint, HBFP}, and recent industry formats like NVFP4 \cite{nvfp4} push bit-widths lower by sharing a single exponent across a block, but when exposed to VLM outliers they suffer from ``microscaling collapse'': the shared exponent is hijacked by the massive outlier, completely erasing background tokens. NVFP4 further requires complex nested scaling hierarchies, coupling block-level FP8 scales with tensor-level FP32 scales, to reach acceptable accuracy. Recent proposals like \textit{MicroScopiQ} \cite{MicroScopiQ} mitigate collapse by borrowing bits from unimportant elements, yet harbor fatal microarchitectural costs: irregular dataflows, deep de-quantization trees, and silicon-heavy multipliers retained inside the PEs, critically diminishing area and power efficiency. The recent outlier-aware FP4 formats MXFP4+ \mbox{\cite{MX_PLUS}} and AMXFP4 \mbox{\cite{amxfp4}} likewise improve accuracy, but enlarge the PE through an extended-mantissa block maximum or two asymmetric FP8 scales.

\begin{figure}[!t]
    \centering
    \includegraphics[width=\linewidth]{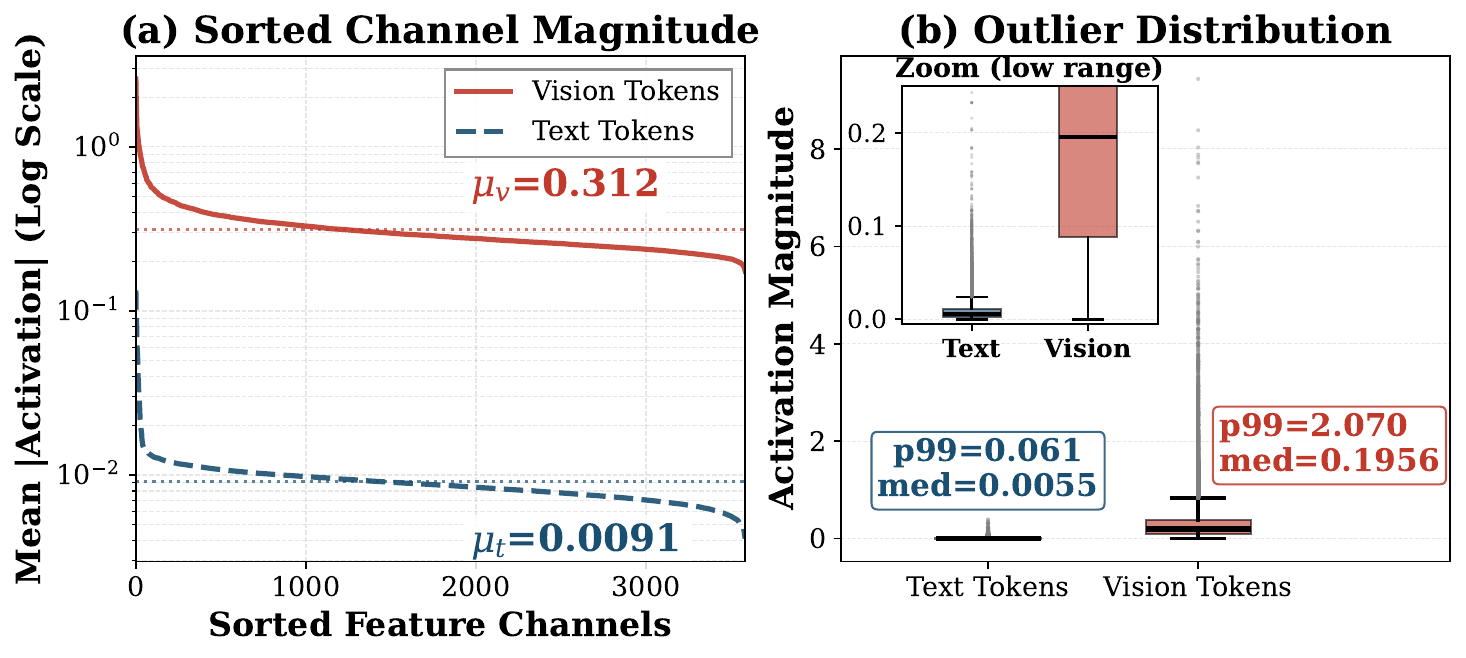}
    \caption{The modality gap between vision and text tokens at the input of the LLM backbone.}
    \label{fig:activation_gap}
    \vspace{-6pt}
\end{figure}

To fundamentally resolve the multi-modal dynamic range bottleneck while aggressively optimizing edge Power, Performance, and Area (PPA), we propose Micro-Inverted-Scaling (\textbf{MiX}), a novel hardware-algorithm co-design framework. MiX mathematically inverts the standard microscaling paradigm: instead of sharing one exponent across multiple mantissas, it groups private, per-element exponents under a single shared mantissa. MiX therefore absorbs extreme intra-block dynamic ranges without destructive clipping, simultaneously preserving visual outliers and fine-grained textual features.

Notably, this algorithmic formulation translates into massive microarchitectural savings. An adaptive dual-format framework pairs a MiX-formatted activation tensor with an MX-formatted weight tensor, allowing the shared MiX mantissa to be algebraically factored out of the block dot product. This simplification lets the custom systolic accelerator eradicate complex MACs entirely, replacing them with efficient bit-shifters and integer adder trees.

To our best of knowledge, MiX is the first end-to-end low-bit VLM accelerator. Evaluated end-to-end on Qwen2-VL \cite{qwen2vl}, LLaVA-OneVision \cite{llava-onevision}, and MiniCPM-V \cite{minicpm-v} across diverse visual-language benchmarks, the MiX-MX architecture dominates the PPA-accuracy Pareto frontier. We implement the MiX accelerator and baseline accelerators in TSMC28 CMOS. A 4.5-bit MiX matches the NVFP4 baseline accuracy while the dedicated MiX accelerator improves PE area efficiency by \mbox{$25\%$} over NVFP4. Against the state-of-the-art accelerator \textit{Focus} \cite{focus}, MiX delivers an average \mbox{$2.3\times$--$4.5\times$} speedup and \mbox{$1.4\times$--$2.9\times$} energy reduction across six VLM benchmarks on LLaVA-OneVision-7B and MiniCPM-V-2.6, proving that the inverted-scaling, multiplier-less datapath is physically superior for end-to-end VLM deployment.

\section{Background and Motivation}

\begin{figure}[!t]
    \centering
    \includegraphics[width=\linewidth]{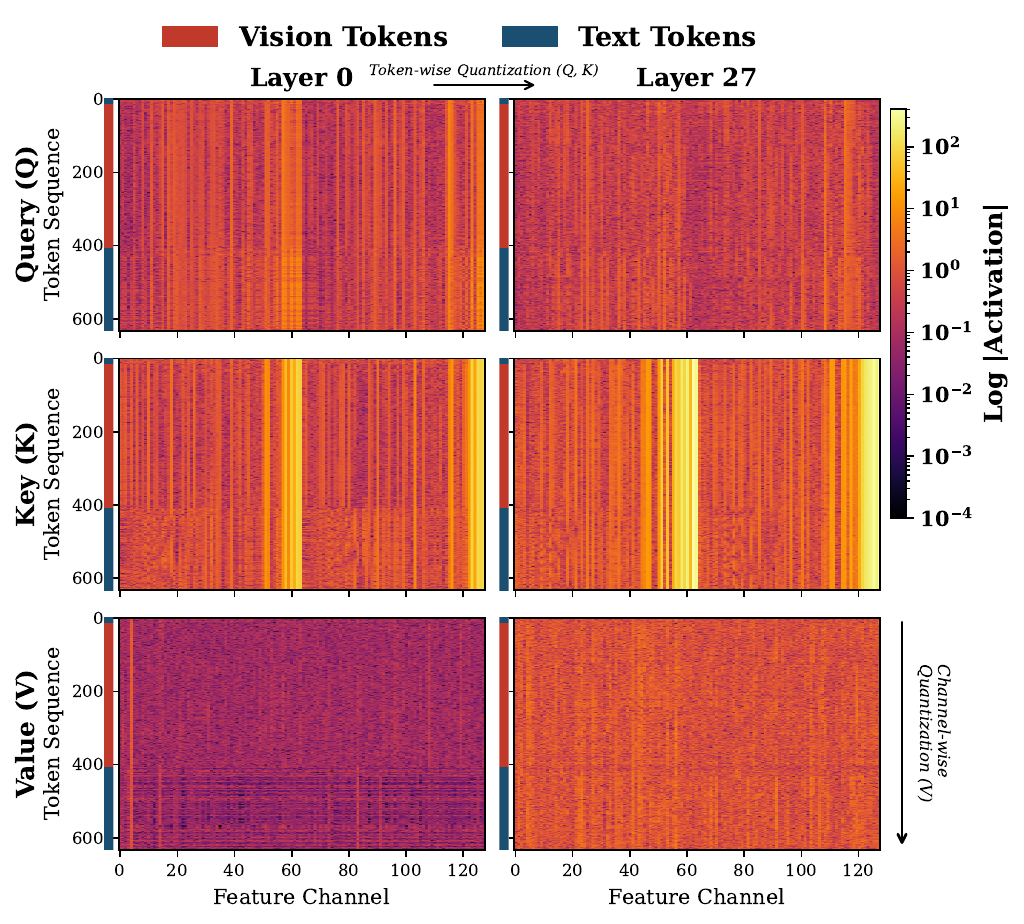}
    \caption{The heatmap of the $Q$, $K$, and $V$ matrices in Qwen2-VL-7B. $Q$ and $K$ exhibit severe channel-wise outliers, while $V$ contains early token-wise spikes.}
    \label{fig:qkv_heatmaps}
\end{figure}

\begin{figure*}[!t]
  \centering
  \includegraphics[width=\textwidth]{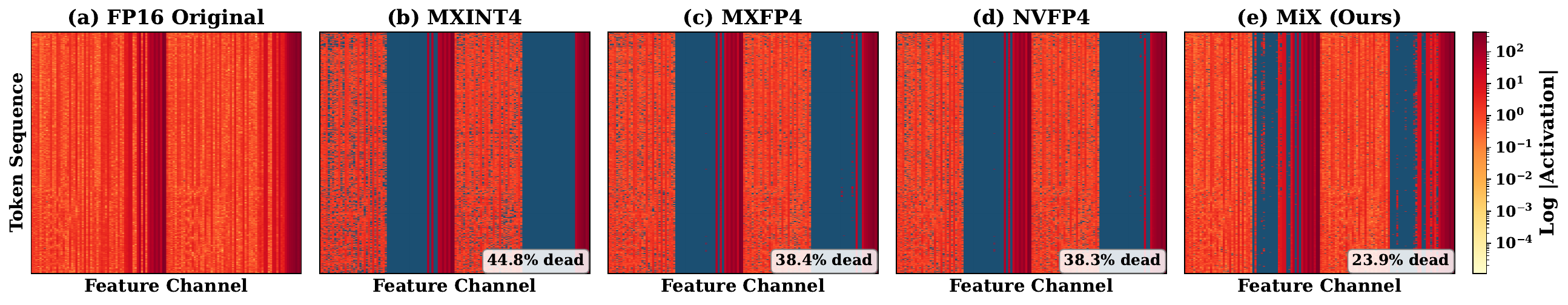}
  \caption{Quantization underflow map of the Layer 27 Key matrix from Qwen2-VL-7B. All formats are in 32 block size.}
  \label{fig:underflow_map}
\end{figure*}

\subsection{VLM Architecture and Multi-Modal Activation Dynamics}
\label{sec:vlm_dynamics}

Modern VLMs process multi-modal inputs through a three-stage pipeline: a Vision Encoder extracts continuous visual patches, a Modality Projector aligns these spatial features into the textual dimension, and an LLM backbone performs the final autoregressive reasoning. Unlike unimodal LLMs, VLMs must dynamically fuse highly diverse modalities, and this fusion begins at the very first layer of the LLM backbone during prefill, where high-density continuous vision tokens are concatenated directly with discrete text embeddings, yielding a highly heterogeneous token sequence. Profiling the hidden states of Qwen2-VL-7B on an OCRBench task (extracting text from a rendered invoice) confirms the hypothesized activation-pattern differences between vision and text tokens.

As \textcolor{refblue}{Figure}~\ref{fig:activation_gap} shows, vision tokens produced by the encoder and projector exhibit a heavily skewed magnitude profile ($\mu_v = 0.312$), whereas text tokens stay tightly concentrated near zero ($\mu_t = 0.0091$), inducing a $\sim$34$\times$ dynamic range gap across feature channels. Element-wise distributions further reveal that vision tokens produce 99th-percentile outliers exceeding 2.07, while text tokens rarely surpass 0.061 at the same percentile.

This initial modality gap further exacerbates extreme outlier channels within the intermediate attention matrices, far more severely than in unimodal LLMs. \textcolor{refblue}{Figure}~\ref{fig:qkv_heatmaps} visualizes the absolute maximum values of the Query ($Q$), Key ($K$), and Value ($V$) matrices across tokens and channels for Qwen2-VL-7B, revealing two distinct outlier behaviors that severely disrupt standard block-wise quantization:
\begin{enumerate}
    \item \textbf{Channel-wise Outliers in $Q$ and $K$:} both matrices exhibit massive, persistent vertical spikes concentrated in specific channels but spanning all tokens, growing with layer depth. 
    \item \textbf{Early Token-wise Spikes in $V$:} the Value matrix instead shows extreme horizontal spikes heavily concentrated in early tokens (compressed and not depicted in the figure).
\end{enumerate}

These intersecting outlier topologies are fatal to standard block-based formats like MX or NVFP4. When a 1D block is mapped along the channel dimension, the massive channel-wise outliers in $Q$ and $K$ force the shared block exponent to scale up dramatically, underflowing the surrounding normal-magnitude background tokens to zero, a phenomenon we term \textit{microscaling collapse}. The result is systematic destruction of attention-map precision and the catastrophic task degradation commonly observed when quantizing VLMs.

\subsection{Block-Quantized Formats}
\label{sec:background_formats}

Block-based low-bit quantization has emerged as the industry standard for alleviating the memory wall in LLM inference. Rather than assigning a unique high-precision scale to every parameter, it groups elements into contiguous chunks (default block size $B=32$) that share a single scale. Microscaling derivatives such as MXINT4 and MXFP4 share a single exponent across the block while keeping individual low-bit mantissas per element. The de-quantized value of the $i$-th element in an MX block is:
\begin{equation}
    x_i = 2^{E_{block}} \times m_i \quad \text{for } i \in \{1, 2, \dots, B\}
\label{eq:mx_format}
\end{equation}
where $E_{block}$ is the shared block exponent derived from the chunk's maximum absolute value, and $m_i$ is the low-bit mantissa in floating-point (MXFP) or integer (MXINT) representation.

More recent formats like NVFP4 ($B=16$) push this paradigm further, but achieving acceptable accuracy in such aggressive sub-8-bit regimes requires complex \textit{nested scaling} hierarchies. NVFP4, for example, multiplies each element by two FP scales:
\begin{equation}
    x_i = S_{tensor} \times S_{block} \times m_i
\label{eq:nvfp4_format}
\end{equation}
where $S_{tensor}$ is a global FP32 scalar applied across the entire matrix or vector and $S_{block}$ is a block-level FP8 scale factor.

\subsection{The Microscaling Collapse in VLMs}
\label{sec:microscaling_collapse}

The \textit{microscaling collapse} is empirically visualized in \textcolor{refblue}{Figure}~\ref{fig:underflow_map}, which analyzes the Layer 27 Key matrix of Qwen2-VL-7B. Under standard shared-scale formats at $B=32$, outlier-dominated scales eradicate non-maximum information within each block, forcing 44.8\% in MXINT4 format and 38.4\% in MXFP4 format of the matrix background tokens into fatal underflow, which are rendered mathematically dead and shown as blue pixels. NVFP4 only slightly mitigates this via a tighter $B=16$ block and a two-level scaling hierarchy. At $B=32$ a 38.3\% underflow rate persists, losing significant multi-modal signal in outlier-contaminated regions.

To rescue such lost representational capacity without raising the nominal bit-width, we introduce Micro-Inverted-Scaling (MiX). By assigning per-element relative exponents, MiX grants local dynamic-range flexibility, allowing background tokens to retain distinct scales even within outlier-heavy blocks. The per-element exponent overhead is amortized by sharing a single block-level mantissa, so MiX preserves a bit budget comparable to conventional MX while redirecting precision toward local range adaptation. At $B=32$, MiX circumvents the structural collapse and reduces total matrix underflow to 23.9\%, marking $>$14\% less dead zone improvement. This algorithm is co-designed with the hardware backend. Since every element in a MiX block carries the same mantissa, the mantissa also factors out of each block dot product, replacing costly per-element multiplications with efficient bit-shifts.

\section{Micro-Inverted-Scaling (MiX) \& Dual-Format}
\label{sec:algorithm}

\subsection{The MiX Quantization Algorithm}
\label{sec:mix_algorithm}

\begin{figure}[!t]
    \centering
    \includegraphics[width=\linewidth,
                    page=1,
                    trim=1 1 1 1,
                    clip]{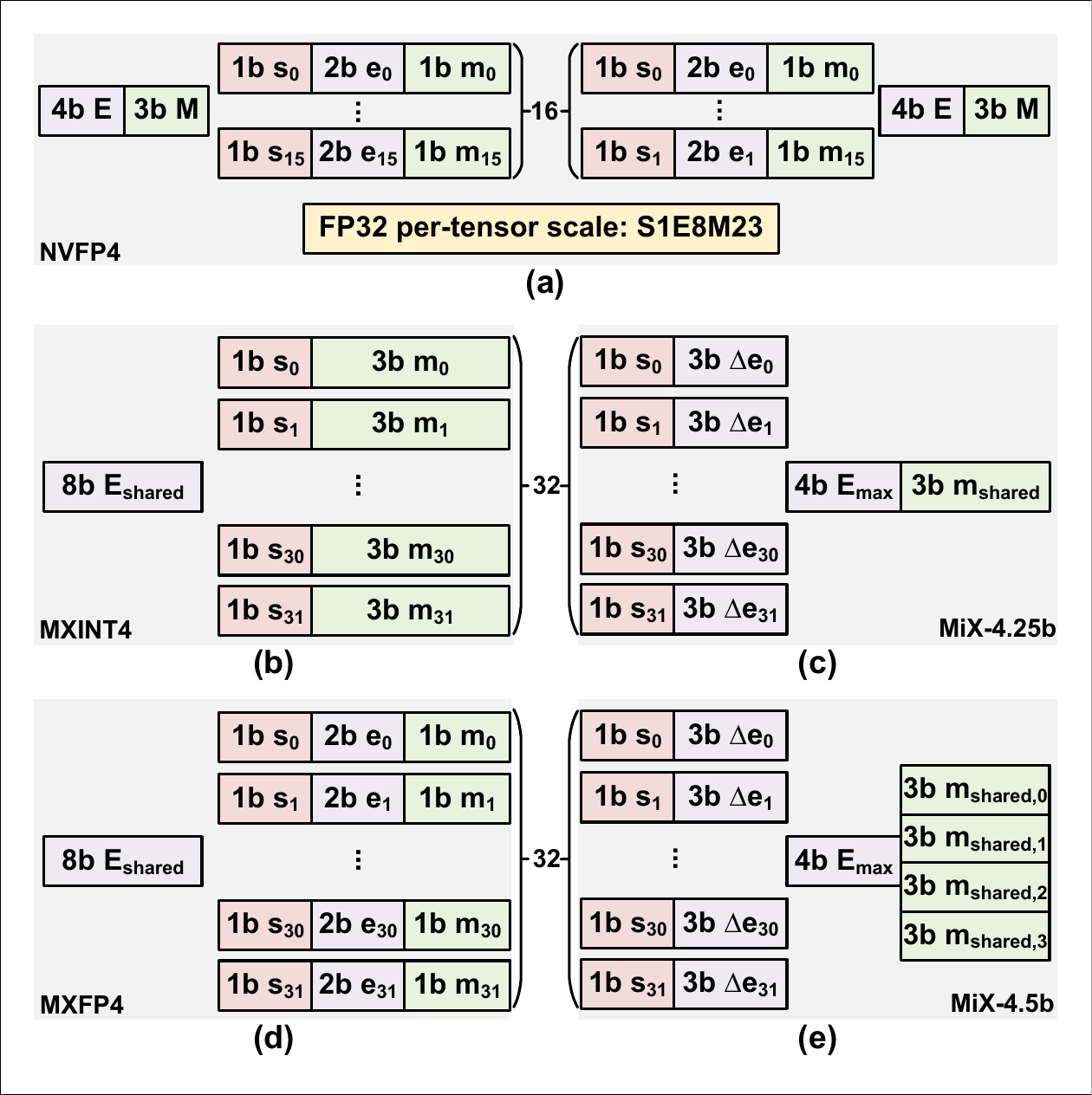}
    \vspace{-20pt}
    \caption{(a) Two NVFP4 blocks and their shared FP32 per-tensor scale. (b) MXINT4 block. (c) MiX-4.25b (Type 1) block. (d) MXFP4 block. (e) MiX-4.5b (Type 2) block.}
    \label{fig:mix_structure}
\end{figure}

Standard microscaling (MX) \cite{microscaling} shares a single exponent across a block to amortize hardware costs, but fundamentally fails when exposed to VLM activation outliers. To break this structural bottleneck, we propose \textbf{Micro-Inverted-Scaling (MiX)}, which mathematically inverts the paradigm: rather than grouping multiple mantissas under a single shared exponent, MiX groups multiple private exponents under one shared mantissa. By allocating the bit-budget primarily to per-element exponent differences, MiX absorbs extreme intra-block dynamic ranges.

To enable rigorous comparisons against state-of-the-art baselines and map a complete Pareto frontier, we formalize MiX into two format variants, each designed to match the effective bit-widths (EBW) of current industry standards.

\vspace{1mm}\noindent\textbf{Type 1: Standard MiX (MiX-4.25b).}
Designed as a bit-for-bit competitor to MXFP4, this baseline operates on a block of size $B=32$, illustrated in \textcolor{refblue}{Figure}~\ref{fig:mix_structure}\textcolor{refblue}{c}. The block is anchored by a single 4-bit maximum base exponent ($E_{max}$) and a 3-bit shared mantissa ($m_{shared}$) with an implicit leading 1. For each element $x_i$, we extract a 1-bit sign ($s_i$) and a 3-bit relative exponent difference ($\Delta e_i$) that dictates the right-shift required to scale the shared mantissa to the element's local magnitude, with the exception that a value of seven encodes zero. The de-quantized value is reconstructed as:
\begin{equation}
    \hat{x}_i = (-1)^{s_i} \times (1.m_{shared})_2 \times 2^{E_{max} - \Delta e_i}
\label{eq:mix_type1}
\end{equation}
Requiring $4+3$ shared bits padded with one spare bit to a regular 8-bit field and $128$ element bits, the EBW is exactly $(8 + 128) / 32 = 4.25$ bits per element, exactly matching the EBW of MXFP4.

\vspace{1mm}\noindent\textbf{Type 2: Sub-Group Mantissas (MiX-4.5b).}
This variant, illustrated in \textcolor{refblue}{Figure}~\ref{fig:mix_structure}\textcolor{refblue}{e}, directly competes with NVFP4's effective footprint. The block ($B=32$) shares a global 4-bit $E_{max}$ but is divided into $K=4$ sub-blocks of size $b=8$, each carrying its own 3-bit shared mantissa ($m_{shared, k}$). The reconstruction becomes:
\begin{equation}
    \hat{x}_{i \in \text{sub}_k} = (-1)^{s_i} \times (1.m_{shared, k})_2 \times 2^{E_{max} - \Delta e_i}
\end{equation}
The bit cost is $4$ (max exponent) $+ 4 \times 3$ (sub-group mantissas) $+ 32 \times 4$ (elements) $= 144$ bits, yielding an EBW of $4.5$ bits per element, matching NVFP4 while eliminating its reliance on tensor-level scaling.

\vspace{1mm}\noindent\textbf{Comparison to MXFP4 and NVFP4.}
\textcolor{refblue}{Figure}~\ref{fig:mix_structure} illustrates the differences between the proposed MiX schemes and MXFP4/NVFP4. MiX-4.25b departs from MXFP4 in two respects. In block scale, MXFP4 carries a single 8-bit power-of-two shared exponent, whereas MiX-4.25b splits the block scale into a 4-bit $E_{max}$ and a 3-bit $m_{shared}$. In element format, MXFP4 stores each element as E2M1, a per-element exponent and mantissa, whereas MiX encodes E3M0, whose three bits are a per-element exponent \emph{difference} $\Delta e$ under the shared mantissa rather than an exponent. Relative to NVFP4, MiX-4.25b differs along four axes: the element format above, block size ($B{=}32$ versus $B{=}16$), scale hierarchy (NVFP4 has an extra S1E8M23 FP32 per-tensor scale), and EBW (MiX-4.25b is a 4.25b format, a quarter-bit below NVFP4's 4.5b). MiX-4.5b reaches NVFP4's 4.5b footprint, so it matches on EBW but still inherits the first three distinctions, and further differs by partitioning each block into four sub-groups with independent shared mantissas.

\begin{algorithm}[!t]
\footnotesize
\caption{MiX-4.5b (Type 2) quantization with exponent refinement.}
\label{alg:mix_quant}
\begin{algorithmic}[1]
\REQUIRE Input activation tensor $\mathbf{X}$, Block size $B=32$, Sub-block size $b=8$.
\ENSURE Quantized and de-quantized tensor $\hat{\mathbf{X}}$.
\FOR{each block $X_{block}$ in $\mathbf{X}$}
    \STATE $E_{max} \leftarrow \max_{x \in X_{block}} \lfloor \log_2(|x|) \rfloor$ \COMMENT{Extract 4-bit max exponent}
    \FOR{$k = 1$ \textbf{to} $B/b$}
        \STATE $X_{sub} \leftarrow X_{block}[(k-1)b : kb]$
        \STATE $m_{shared, k} \leftarrow \text{QuantizeMantissa}(\max(|X_{sub}|), \text{bits}=3)$
        \FOR{each $x_i \in X_{sub}$}
            \STATE $s_i \leftarrow \text{Sign}(x_i)$
            \STATE $\Delta e_{naive} \leftarrow \min(E_{max} - \lfloor \log_2(|x_i|) \rfloor, 7)$
            \STATE $\mathcal{S} \leftarrow \{ \max(0, \Delta e_{naive}-1), \Delta e_{naive}, \min(7, \Delta e_{naive}+1) \}$
            \STATE \COMMENT{EXP Refine: Search only the $\pm 1$ neighborhood}
            \STATE $\Delta e_i \leftarrow \arg\min_{\delta \in \mathcal{S}} \left| x_i - (-1)^{s_i} (1.m_{shared, k})_2 2^{E_{max} - \delta} \right|$
            \STATE $\hat{x}_i \leftarrow (-1)^{s_i} \times (1.m_{shared, k})_2 \times 2^{E_{max} - \Delta e_i}$
        \ENDFOR
    \ENDFOR
\ENDFOR
\end{algorithmic}
\end{algorithm}

\vspace{1mm}\noindent\textbf{Algorithmic Implementation and Exponent Finetuning.}
The baseline MiX quantization process, detailed in \textcolor{refblue}{Algorithm}~\ref{alg:mix_quant}, begins by identifying the maximum absolute value within the block to extract the global $E_{max}$ and shared mantissa $m_{shared}$. We set $m_{shared}$ to the mantissa of the block maximum, which outperforms mean-, scale-weighted, and MSE-based alternatives while saving significant hardware complexity. Selecting maximum's mantissa preserves the important outliers magnitude, maintaining the best accuracy in the experiments of validating all alternatives. However, since the mantissa is globally fixed for the surrounding elements, na\"ively assigning the relative exponent ($\Delta e_i$) from the logarithmic difference often introduces more quantization noise.

To minimize the local reconstruction error $|\hat{x}_i - x_i|$, localized \textit{exponent finetuning} is performed (\textcolor{refblue}{Algorithm}~\ref{alg:mix_quant}, Lines 9--11), which computes the 
exponent difference and searches the $\pm 1$ neighbor for the candidate that minimizes reconstruction error. While trivial in software calibration, this finetuning is also implemented in our hardware accelerator. As detailed in \textcolor{refblue}{Section}~\ref{sec:microarch}, an efficient quantizer at the output datapath resolves the optimal $\Delta e_i$ across the 3-candidate search space through bit-wise logic alone, eliminating any software-emulation overhead.

\subsection{MiX-MX Matrix Multiplication}
\label{sec:dual_format_gemm}

To execute VLM inference efficiently, we propose an adaptive, dual-format matrix multiplication strategy. We formalize the dual-format dot product of MiX-format activation (A-MiX) $\times$ MX-format weight (W-MX). 
Consider a dot product between A-MiX block $\mathbf{x}$ (with MiX-4.25b Type 1 scheme) and a W-MX block $\mathbf{w}$, both of size $B=32$ (\textcolor{refblue}{Figure}~\ref{fig:mix_structure}\textcolor{refblue}{b} and~\ref{fig:mix_structure}\textcolor{refblue}{c}). Following the definitions from \textcolor{refblue}{Section}~\ref{sec:mix_algorithm}, the exact block dot product $y = \sum_{i=1}^{B} \hat{x}_i \hat{w}_i$ expands as:
\begin{equation}
\begin{split}
    y = \sum_{i=1}^{B} & \left( (-1)^{s_{x,i}} (1.m_{shared})_2\, 2^{E_{max} - \Delta e_{x,i}} \right) \\
                       & \times \left( (-1)^{s_{w,i}} m_{w,i}\, 2^{E_w} \right)
\end{split}
\end{equation}

Standard quantization algorithms require an $m_{x,i} \times m_{w,i}$ multiplication for every element pair. However, because the MiX format groups multiple private exponents under one shared mantissa as shown in \textcolor{refblue}{Figure}~\ref{fig:mix_structure}\textcolor{refblue}{c}, the $(1.m_{shared})_2$ term and the global base exponents can be factored out of the summation entirely:
\begin{equation}
\begin{split}
    y = {} & \underbrace{(1.m_{shared})_2}_{\text{Scalar Multiply}}
             \times \underbrace{2^{E_{max} + E_w}}_{\text{Exponent Addition}} \\
           & \times \sum_{i=1}^{B} \underbrace{\left[ (-1)^{s_{x,i} \oplus s_{w,i}}
             \left( m_{w,i} \times 2^{-\Delta e_{x,i}} \right) \right]}_{\text{Shift-and-Add Logic}}
\end{split}
\label{eq:mix_mx_dot_product}
\end{equation}

This factorization reduces the element-wise operations to $m_{w,i} \times 2^{-\Delta e_{x,i}}$, which maps directly to a right-shift of the weight mantissa by the activation's relative exponent.
Consequently, the algorithmic steps required to compute the block dot product become massively simplified:
\begin{enumerate}
    \item \textbf{Shift:} Shift the MX weight mantissas by the per-element MiX relative exponents ($\Delta e_{x,i}$).
    \item \textbf{Accumulate:} Sum the shifted mantissas to form a single intra-block partial sum.
    \item \textbf{Scale:} Apply exactly \textit{one} multiplication to scale the accumulated sum by the shared $(1.m_{shared})_2$ activation mantissa, while adding the globally shared base exponents ($E_{max} + E_w$) to the exponent block.
\end{enumerate}

By leveraging this dual-format algebra, the MiX-MX co-design eliminates per-element multiplications, paving the way for a multiplier-less accelerator architecture.

\subsection{Signal-to-Quantization-Noise Analysis}

\begin{figure}[!t]
    \centering
    \includegraphics[width=\linewidth]{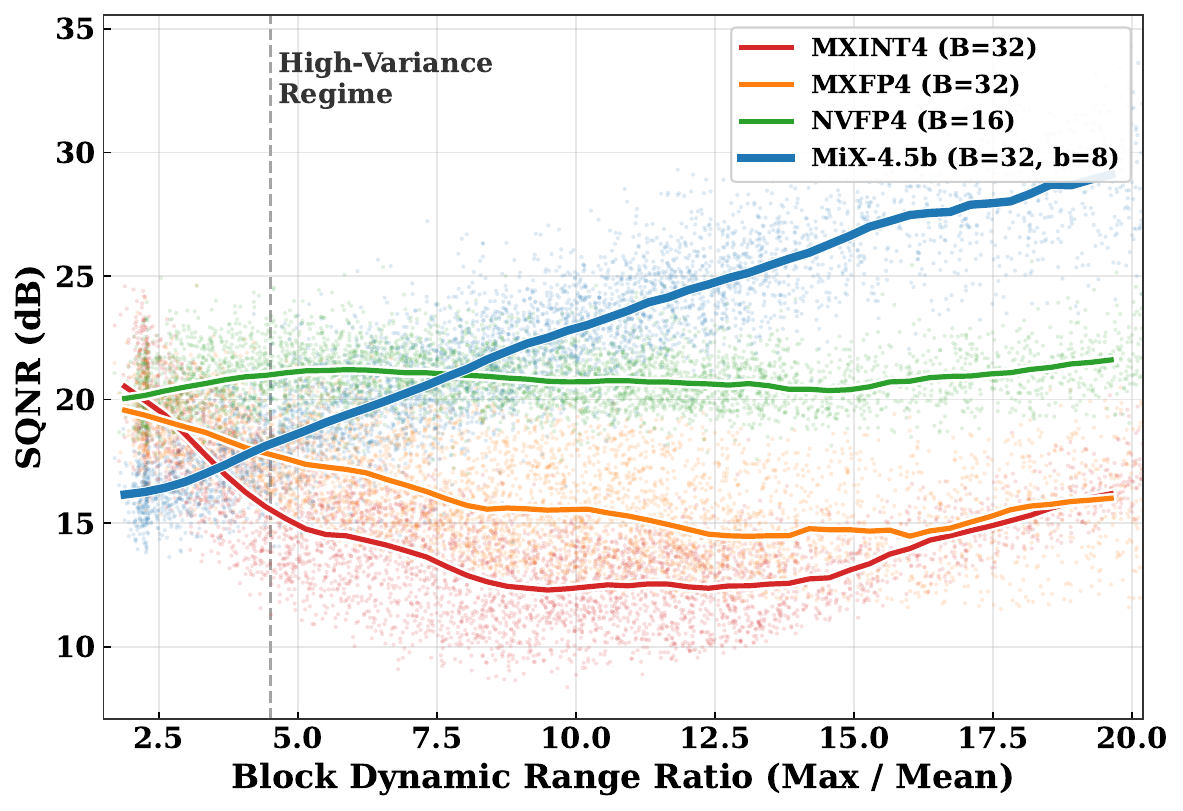}
    \caption{Per-block Signal-to-Quantization-Noise Ratio across varying intra-block dynamic ranges for Qwen2-VL-7B hidden-state activations.}
    \label{fig:sqnr_analysis}
    \vspace{\fpad}
\end{figure}

To quantify the representational fidelity of MiX against baseline formats, we analyze the per-block Signal-to-Quantization-Noise Ratio (SQNR), defined as the ratio of signal power ($\sum x^2$) to quantization noise power ($\sum (x-\hat{x})^2$), on Layer 22 hidden-state activations from Qwen2-VL-7B. In \textcolor{refblue}{Figure}~\ref{fig:sqnr_analysis}, each plotted point corresponds to one 32-element block, the horizontal axis measures the intra-block dynamic-range (DR) ratio (block maximum divided by block mean absolute value), and the vertical axis reports the SQNR after a full quantize--dequantize cycle.

In the low-variance regime (DR ratio below the dashed boundary at 4.5), all evaluated formats start from a comparable baseline fidelity between 16 and 21 decibels, since a single shared scale adequately represents every element in the block. Past that boundary, however, the shared-scale formats suffer a pronounced collapse, MXINT4 bottoming out near 12 dB and MXFP4 near 14.5 dB: their shared exponent is pinned to the block maximum, forcing smaller elements into coarser quantization levels or outright underflow. NVFP4 mitigates this with a smaller block size of 16 but plateaus near 21 dB as its block scale hits its precision limit. The non-monotonic recovery of MXINT4 and MXFP4 beyond a ratio of roughly 13 is an artifact of the power-weighted noise metric rather than a true precision recovery: once an outlier becomes extreme, the underflowed elements are numerically negligible relative to the total signal power, masking the underlying dead-zone failure.

In contrast, the proposed MiX format exhibits a steadily rising fidelity curve, surpassing all baselines beyond DR ratio of 8 and reaching approximately 28 dB at DR ratio of 18. Its per-element exponents independently track each element's magnitude while the shared mantissa within each 8-element sub-group preserves relative precision. This resilience to high intra-block DR makes MiX particularly well-suited for VLM activations, where the 34$\times$ vision--text magnitude gap from \textcolor{refblue}{Figure}~\ref{fig:activation_gap} precisely produces such extreme intra-block variance.

\section{Architecture: Multiplier-Less Systolic Array}
\label{sec:microarch}

To physically instantiate the algorithmic factorization and adaptive dual-format mapping established in \textcolor{refblue}{Section}~\ref{sec:algorithm}, we design a specialized, multiplier-less spatial accelerator. The proposed architecture executes MiX-MX matrix multiplications while minimizing data movement.



\begin{figure}[!t]
    \centering
    \includegraphics[width=0.75\linewidth,
                    page=2,
                    trim=1 1 1 1,
                    clip]{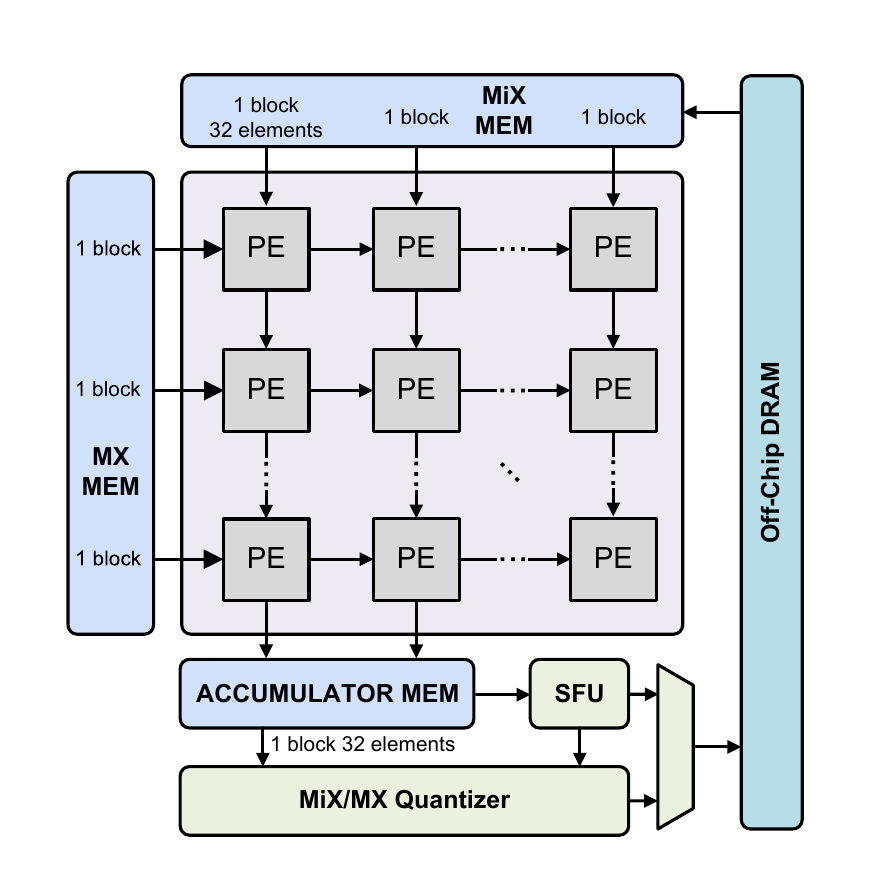}
    \caption{Top-level architecture of the accelerator.}
    \label{fig:top_arch}
\end{figure}

\subsection{Overall Architecture and Dataflow}
\label{sec:overall_arch}

\textcolor{refblue}{Figure}~\ref{fig:top_arch} shows the top-level architecture of the proposed accelerator. To support A-MiX/W-MX matrix multiplications, weight and activation buffers are designated as \textbf{MX Memory} and \textbf{MiX Memory}, respectively, feeding the horizontal and vertical inputs of the systolic array.

The computational core is a 2D Systolic Array executing an Output-Stationary (OS) dataflow. Unlike standard systolic arrays that process individual elements per cycle, the MiX-MX factorization in \textcolor{refblue}{Equation~\eqref{eq:mix_mx_dot_product}} lets each PE ingest an entire quantization block in a single cycle.

Upon completing a matrix tile, the stationary high-precision results are shifted downward along the vertical accumulation chain into the \textbf{Accumulator Memory} and then routed through a specialized \textbf{MiX/MX Quantizer} or Special Function Unit (SFU) at the output datapath. The quantizer is the hardware realization of \textcolor{refblue}{Algorithm}~\ref{alg:mix_quant}'s \textit{exponent refinement}, dynamically re-compressing the activations into the required MiX or MX format for the subsequent layer. The result of the quantizer or SFU is saved to Off-Chip DRAM.

\begin{figure}[!t]
    \centering
    \includegraphics[width=\linewidth,
                    page=3,
                    trim=1 1 1 1,
                    clip]{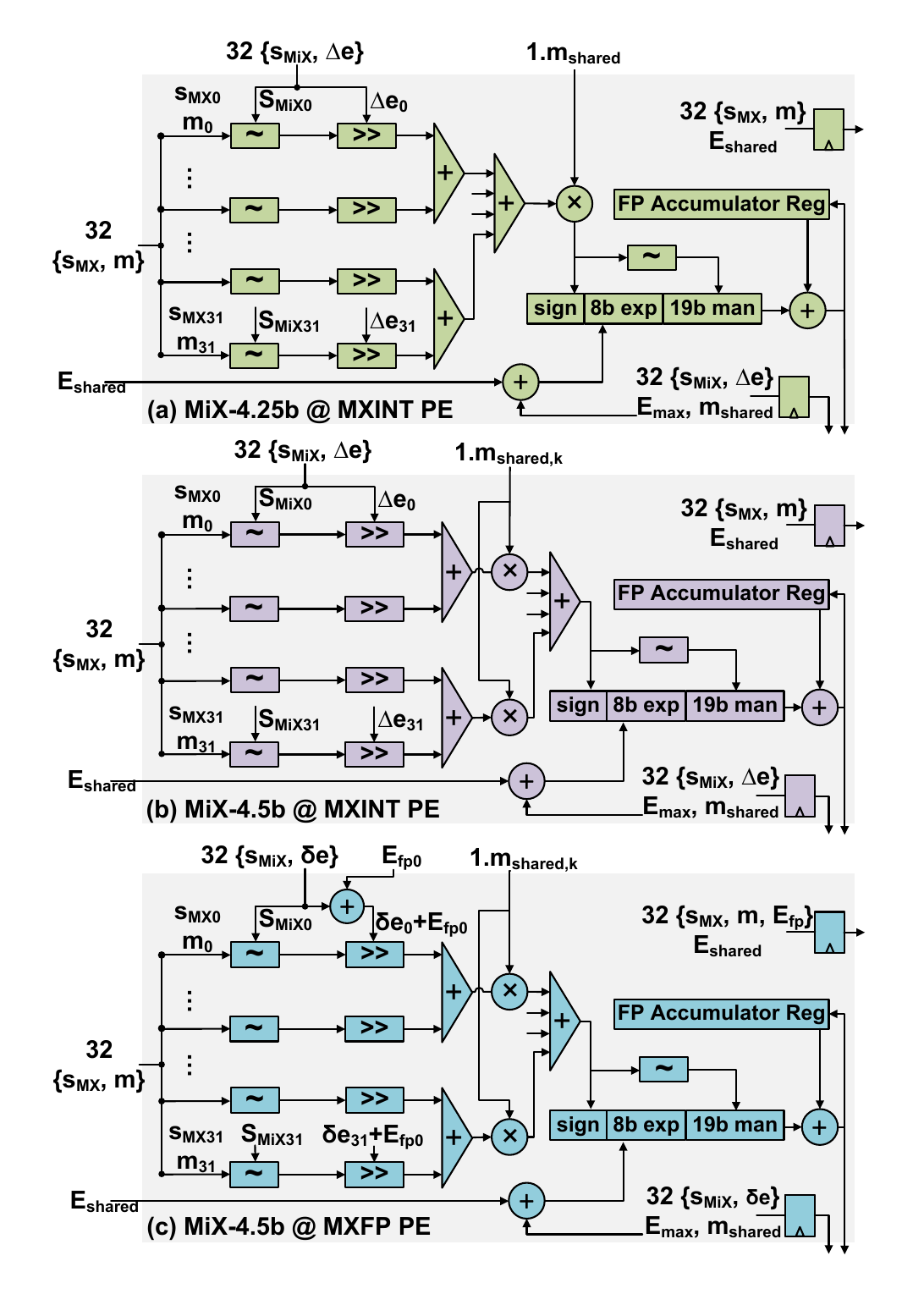}
    \caption{The PE microarchitecture and intra-PE shift-and-add logic for the different MiX format variants.}
    \label{fig:pe_arch}
\vspace{\fpad}
\end{figure}

\subsection{PE Microarchitecture}
\label{sec:pe_arch}

At the core of the spatial accelerator is the multiplier-less PE, which executes the algebraic factorization of \textcolor{refblue}{Equation~\eqref{eq:mix_mx_dot_product}}, ingesting 32 parallel MiX elements and 32 parallel MX elements per cycle. As \textcolor{refblue}{Figure}~\ref{fig:pe_arch} illustrates, the microarchitecture adapts to the format variant. For clarity, we first describe the baseline MiX-4.25b datapath before detailing the structural adaptations for the sub-group variants.

\vspace{1mm}\noindent\textbf{The Shift-and-Add Datapath.} The intra-PE computation operates entirely in the integer domain prior to final accumulation, proceeding through the following stages:

\textit{Sign Processing:} The MXINT integer is conditionally inverted using the sign bit of the corresponding MiX element, realizing the $(-1)^{s_{x,i} \oplus s_{w,i}}$ term.

\textit{Shifting:} The signed MXINT integer enters a barrel shifter whose shift amount is the MiX element's 3-bit relative exponent ($\Delta e_i$), executing a logical right-shift that aligns the integer with the MiX element's magnitude.

\textit{Integer Adder Tree:} The 32 shifted integers feed a parallel integer adder tree that produces a single unscaled partial sum for the block.

\vspace{1mm}\noindent\textbf{Scalar Multiplication and FP Accumulation.} To complete the factorization, the unscaled partial sum is multiplied by the 4-bit shared MiX mantissa, where the explicit 3 bits are concatenated with the implicit leading 1.

The hardware then transitions from the integer to the FP domain. The scalar product becomes the mantissa and sign of a temporary FP variable. The internal 19-bit mantissa datapath is wide enough to hold every precision bit produced by the adder tree and multiplier without immediate truncation. The exponent is constructed by adding the MiX block's maximum base exponent ($E_{max}$) to the MX block's shared exponent ($E_{shared}$). This constructed FP number is fed into a standard FP32 adder that updates the PE's local accumulation register. Simultaneously, the input MiX and MX blocks are captured in pipeline flip-flops and forwarded to adjacent PEs to sustain the spatial wavefront.

\vspace{1mm}\noindent\textbf{Microarchitectural Adaptations for Sub-Group Variants.} The PE datapath is modular, adapting to the finer MiX-4.5b format and the MiX-MXFP dual-format without fundamentally altering the pipeline, shown in bottom two PEs depicted in \textcolor{refblue}{Figure}~\ref{fig:pe_arch}.

\textit{MiX-4.5b @ MXINT:} To support the four sub-groups of size $b=8$, the 32-wide integer adder tree is partitioned into four independent 8-wide sub-trees. Each sub-tree feeds its own scalar multiplier that multiplies the sub-sum by the corresponding sub-group mantissa ($m_{shared, k}$), and the four products are merged by a secondary adder stage before final FP conversion.

\textit{MiX-4.5b @ MXFP:} The MXFP variant builds directly upon the MiX-4.5b @ MXINT datapath, adding one low-cost pre-processing step. Before the barrel shifters, a narrow 3-bit adder sums the FP4 exponent with the element's local difference ($\delta e_i$) to produce the total shift amount, which then drives the existing shift-and-add logic.

\subsection{Dual-Format Quantizer}
\label{sec:quantizer}

\begin{figure}[!t]
    \centering
    \includegraphics[width=\linewidth,
                    page=4,
                    trim=3 1 1 1,
                    clip]{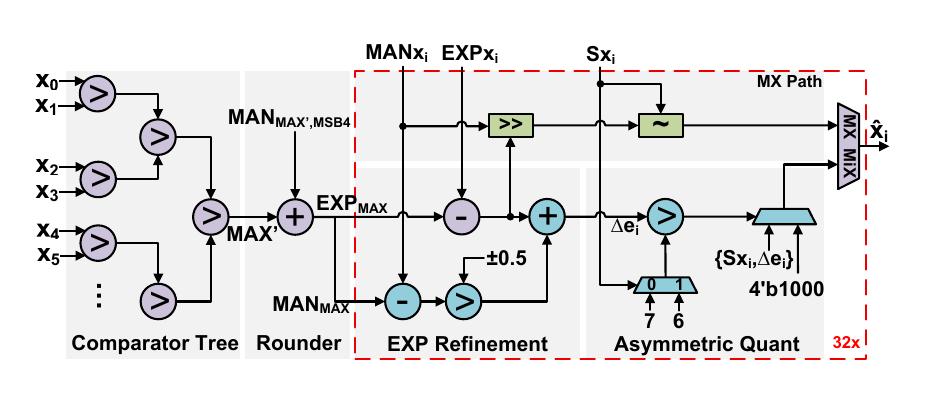}
    \caption{Microarchitecture of the fused MiX/MX output quantizer. Purple for shared path, blue for MiX only path and green for MX only path.}
    \label{fig:quantizer}
\end{figure}

The systolic array emits FP32 accumulations. To sustain the adaptive dual-format inference strategy across sequential layers, these temporary FP values must be re-compressed before being written back to global SRAM. We design a fused, single-cycle MiX/MX hardware quantizer at the output datapath, as shown in \textcolor{refblue}{Figure}~\ref{fig:quantizer}.

The quantizer maximizes area efficiency by sharing a common feature-extraction frontend (purple components in \textcolor{refblue}{Figure}~\ref{fig:quantizer}). The 32 uncompressed FP values enter a parallel comparator tree that isolates the maximum absolute value of the block. The hardware then extracts its sign, exponent, and 4 Most Significant Bits (MSBs) of mantissa, establishing the baseline metadata shared by both formats. The datapath then diverges based on the target format.

\vspace{1mm}\noindent\textbf{The MiX Datapath.}
When targeting the MiX format, the logic extracts the shared mantissa and computes the optimized relative exponents $\Delta e_i$ across three specialized sub-modules.

\textit{Carry-Propagated Rounding:} To compute the 3-bit shared mantissa $m_{shared}$, the extracted 4-bit mantissa undergoes round-to-nearest logic by adding its 4th bit to the upper 3 MSBs. The hardware implicitly handles carry propagation: if rounding overflows the mantissa (e.g., $111_2 + 1_2 = 1000_2$), the overflow bit routes directly into the exponent adder, incrementing $EXP_{max}$ and leaving $000_2$ as the normalized mantissa.

\textit{Hardware Exponent Refinement:} This module is the physical realization of the error-minimization search in \textcolor{refblue}{Algorithm}~\ref{alg:mix_quant}. Instead of an iterative floating-point search, the logic leverages a single 4-bit mantissa subtraction. Because adjusting a base-2 exponent by 1 scales the reconstructed value by a factor of $2$, the decision boundary that minimizes absolute error between two adjacent quantization bins lies exactly at the $0.5$ midpoint in the linear mantissa space. The hardware computes the residual between the local element's de-normalized mantissa and the rounded $MAN_{max}$. A simple comparator evaluates this residual against the $0.5$ threshold, and the EXP Refine block deterministically emits an adjustment of $+1$, $-1$, or $0$. This adjustment is applied to the naive relative shift, reducing the algorithmic finetuning to a single-cycle, low-latency thresholding operation.

\textit{Asymmetric Zero-Encoding:} To maximize representational density, the quantizer adopts an asymmetric state encoding. Standard sign-magnitude formats waste a state on $-0$ alongside $+0$. We reclaim the redundant negative zero to encode an extra dynamic-range level for positive numbers. Positive numbers therefore evaluate $\Delta e_i$ against a maximum threshold of $7$ while negative numbers use $6$, and values exceeding these bounds are multiplexed to zero.

\vspace{1mm}\noindent\textbf{The MX Datapath.}
When the block targets a standard MX format (purple and green path in \textcolor{refblue}{Figure}~\ref{fig:quantizer}), the comparator tree and the rounder is shared. The shared block exponent is fixed to $EXP_{max}$. The hardware computes each element's shift amount via a parallel subtraction ($EXP_{max} - EXP_{x_i}$), and the FP mantissa is shifted and then conditionally inverted into a signed integer representation. A final multiplexer selects between the MiX and MX vectors, allowing the accelerator to alternate data formats on the fly.

\section{Evaluation Methodology}
\label{sec:methodology}

To comprehensively evaluate the MiX co-design framework, we establish an end-to-end pipeline spanning multi-modal task accuracy, RTL-level microarchitectural efficiency, and system-level performance.

\subsection{Models and Workloads}
We target the edge deployment of state-of-the-art VLMs with 7B--8B parameters, which represent the critical edge-deployment threshold: footprints are large enough to make memory bandwidth a severe bottleneck, yet representational capacity remains strong enough for complex real-world applications. 
Our evaluation suite covers Qwen2-VL-7B \cite{qwen2vl} and LLaVA-OneVision-7B \cite{llava-onevision} for direct comparison with state-of-the-art VLM quantization work, plus the 8B model MiniCPM-V-2.6 \cite{minicpm-v} as an additional target. This 7B to 8B range is the practical sweet spot for edge deployment, efficient to serve yet strong on real tasks. To probe scale and modality, we further evaluate Qwen2.5-VL from 3B to 72B and text-only LLMs.

End-to-end task accuracy is assessed across six multi-modal benchmarks chosen to stress different activation topologies: logical reasoning (MMMU \cite{MMMU}); text-rich visual comprehension (OCRBench \cite{OCRBench}, TextVQA \cite{TextVQA}, ChartQA \cite{ChartQA}); and general perception (VizWiz \cite{VizWiz}, SEED-Bench-2+ \cite{SEEDBench2Plus}). In \textcolor{refblue}{Table}~\ref{tab:accuracy_results}, OCR denotes OCRBench, VW VizWiz, TQA TextVQA, CQA ChartQA, and SEED2 SEED-Bench-2+.

\subsection{Algorithm and Hardware Baselines}

\vspace{1mm}\noindent\textbf{Algorithm Baselines and End-to-End Coverage.}
We benchmark MiX-MX against state-of-the-art VLM Post-Training Quantization (PTQ) algorithms, MBQ \cite{MBQ} and TLQ \cite{TLQ}. As summarized in \textcolor{refblue}{Table}~\ref{tab:quant_coverage}, existing PTQ methods restrict quantization to the linear layers, leaving the Modality Projector and the Attention BMMs in both the Vision Encoder and the LLM Backbone in FP16. In contrast, MiX is applied strictly end-to-end, ensuring that the entire execution pipeline can be offloaded to a sub-8-bit edge accelerator.

\vspace{1mm}\noindent\textbf{Hardware Format Baselines.}
We evaluate MiX against sub-8-bit industry standards: the OCP-standard MXFP4 \cite{microscaling}, its derivative MXINT4, and Nvidia's NVFP4. Because NVFP4 uses a block size of 16 and an EBW of 4.5b, we additionally extend the two MX formats to MXFP4$_{g16}$ and MXINT4$_{g16}$. The corresponding MiX variants are paired with each MX format for direct comparison: MiX-INT4 pairs MiX-4.25b with MXINT4; MiX-FP4 pairs MiX-4.25b with MXFP4, respectively, while MiX-INT4$_{g16}$ and MiX-FP4$_{g16}$ pair MiX-4.5b with MXINT4$_{g16}$ and MXFP4$_{g16}$. We group all formats into two tiers by EBW: 4.25b and 4.5b. We further compare against two recent outlier-aware 4.5b FP4 baselines: \mbox{MXFP4+~\cite{MX_PLUS}}, which repurposes the block-maximum exponent field as extended mantissa, and \mbox{AMXFP4~\cite{amxfp4}}, which applies two asymmetric FP8 block scales.

\begin{table}[!t]
    \centering
    \caption{Quantization coverage across VLM components. Attention BMMs include KV cache quantization.}
    \label{tab:quant_coverage}
    \resizebox{\linewidth}{!}{
    \begin{tabular}{l c c c c c}
        \toprule
        \textbf{Method} & \multicolumn{2}{c}{\textbf{Vision Encoder}} & \textbf{Projector} & \multicolumn{2}{c}{\textbf{LLM Backbone}} \\
        \cmidrule(lr){2-3} \cmidrule(lr){5-6}
        & \textbf{Linear} & \textbf{Attention BMMs} & & \textbf{Linear} & \textbf{Attention BMMs} \\
        \midrule
        TLQ \cite{TLQ} & W/A & \xmark & \xmark & W/A & \xmark \\
        MBQ \cite{MBQ}    & W/A & \xmark & \xmark & W/A & \xmark \\
        Q-VLM \cite{Q-VLM}  & W/A & \xmark & \xmark & W/A & \xmark \\
        VEQ \cite{VEQ}    & \xmark & \xmark & \xmark & W only & \xmark \\
        \midrule
        \textbf{MiX-MX (Ours)} & W/A & W/A & W/A & W/A & W/A \\
        \bottomrule
    \end{tabular}
    }
\end{table}

\vspace{1mm}\noindent\textbf{Accelerator Baseline.}
The recently proposed VLM accelerator \textit{Focus} \cite{focus} executes on a $32 \times 32$ FP16 systolic array and achieves its speedup through visual-token sparsity exploitation. We invoke the Focus simulator \cite{focus} to obtain its speedup gains across the six benchmarks and overlay these results onto our implemented FP16 systolic baseline, enabling a fair comparison between MiX, vanilla FP16 execution, and the algorithmically optimized Focus accelerator.

\subsection{Accelerator Specifications}
To ensure fair PPA comparisons, we evaluate every data format on an iso-throughput PE and systolic-array design. All formats are instanced identically as a \mbox{$4\times4$} array of 32-wide-ingestion PEs delivering 512 MACs per cycle, so every design sustains the same throughput and differs only in its per-format PE datapath. Accelerators for tier 4.5b format takes in 2 blocks per PE to maintain the same 32 MACs per cycle per PE specification as tier 4.25b. The sole exception is the FP16 baseline, which retains a conventional \mbox{$32\times32$} multiply-accumulate array normalized to the same 512-MAC throughput.

\begin{table}[htbp]
    \centering
    \caption{End-to-end VLM task accuracy. * denotes the results adopted from prior works.}
    \label{tab:accuracy_results}
    \renewcommand{\arraystretch}{1.18}%
    \resizebox{\columnwidth}{!}{
    \begin{tabular}{l c c c c c c | c}
        \toprule
        \textbf{Method} & \textbf{MMMU} & \textbf{OCR} & \textbf{VW} & \textbf{TQA} & \textbf{CQA} & \textbf{SEED2} & \textbf{Avg.} \\
        \midrule
        \multicolumn{8}{c}{\textbf{LLaVA-OneVision-7B}} \\
        \midrule
        FP16 & 49.6 & 62.6 & 59.8 & 77.2 & 80.2 & 65.0 & 65.7 \\
        \cmidrule{1-8}
        \multicolumn{8}{c}{\textit{W4A6 / Sparsity-aware / 4.88b}} \\
        MBQ* & 40.6 & 48.1 & 56.3 & 65.4 & 71.5 & 60.0 & 57.0  \\
        TLQ* & 41.4 & 49.8 & 56.3 & 65.9 & 73.0 & 60.7 & 57.9  \\
        Focus & 48.2 & 34.1 & 59.8 & 67.5 & 59.0 & 61.2 & 55.0 \\
        \rowcolor[HTML]{ECF4FF} \textbf{MiX-INT5} & 44.4 & 57.5 & 58.0 & 74.5 & 78.1 & 62.2 & 62.4 \\
        \cmidrule{1-8}
        \multicolumn{8}{c}{\textit{4.50b}} \\
        MXFP4$_{g16}$ & 40.0 & 48.6 & 61.3 & 65.3 & 71.3 & 56.4 & 57.1 \\
        \rowcolor[HTML]{ECF4FF} \textbf{MiX-FP4$_{g16}$} & 42.8 & 53.4 & 61.4 & 68.4 & 73.8 & 59.4 & 59.9 \\
        MXINT4$_{g16}$ & 43.8 & 54.3 & 58.2 & 68.8 & 74.6 & 60.3 & 60.0 \\
        \rowcolor[HTML]{ECF4FF} \textbf{MiX-INT4$_{g16}$} & 47.2 & 56.1 & 58.3 & 70.2 & 76.8 & 61.2 & 61.6 \\
        NVFP4 & 45.9 & 55.3 & 55.8 & 71.1 & 75.6 & 62.3 & 61.0 \\
        MXFP4+ & 46.1 & 57.6 & 58.2 & 71.9 & 76.5 & 62.8 & 62.2 \\
        AMXFP4 & 46.3 & 59.9 & 60.1 & 72.4 & 77.1 & 62.0 & 63.0 \\
        \cmidrule{1-8}
        \multicolumn{8}{c}{\textit{4.25b}} \\
        MXFP4 & 40.2 & 45.8 & 61.0 & 63.4 & 70.5 & 56.1 & 56.2 \\
        \rowcolor[HTML]{ECF4FF} \textbf{MiX-FP4} & 42.1 & 48.5 & 61.3 & 64.6 & 72.0 & 58.8 & 57.9 \\
        MXINT4 & 38.8 & 43.9 & 58.1 & 61.6 & 70.3 & 55.3 & 54.7 \\
        \rowcolor[HTML]{ECF4FF} \textbf{MiX-INT4} & 42.6 & 50.8 & 57.5 & 65.9 & 72.9 & 60.5 & 58.4 \\
        \midrule
        \multicolumn{8}{c}{\textbf{Qwen2-VL-7B}} \\
        \midrule
        FP16 & 50.4 & 83.9 & 66.2 & 84.8 & 83.2 & 69.3 & 73.0 \\
        \cmidrule{1-8}
        \multicolumn{8}{c}{\textit{W4A6 / 4.88b}} \\
        MBQ* & 41.1 & 65.8 & 49.8 & 70.0 & 70.8 & 62.3 & 60.0 \\
        TLQ* & 41.9 & 68.4 & 52.9 & 69.4 & 72.2 & 64.3 & 61.5 \\
        \rowcolor[HTML]{ECF4FF} \textbf{MiX-INT5} & 47.6 & 81.5 & 65.3 & 83.2 & 80.8 & 66.1 & 70.8 \\
        \cmidrule{1-8}
        \multicolumn{8}{c}{\textit{4.50b}} \\
        MXFP4$_{g16}$ & 44.7 & 74.5 & 63.3 & 79.1 & 76.5 & 62.5 & 66.8 \\
        \rowcolor[HTML]{ECF4FF} \textbf{MiX-FP4$_{g16}$} & 44.7 & 76.6 & 65.4 & 80.5 & 78.9 & 64.6 & 68.4 \\
        MXINT4$_{g16}$ & 43.0 & 74.9 & 63.9 & 79.9 & 78.4 & 63.2 & 67.2 \\
        \rowcolor[HTML]{ECF4FF} \textbf{MiX-INT4$_{g16}$} & 46.3 & 78.1 & 64.8 & 81.0 & 80.5 & 65.1 & 69.3 \\
        NVFP4 & 45.0 & 78.9 & 68.1 & 81.6 & 79.2 & 65.2 & 69.7 \\
        MXFP4+ & 45.1 & 78.3 & 66.4 & 82.3 & 80.5 & 66.1 & 69.8 \\
        AMXFP4 & 46.8 & 80.0 & 65.4 & 82.8 & 81.6 & 66.4 & 70.5 \\
        \cmidrule{1-8}
        \multicolumn{8}{c}{\textit{4.25b}} \\
        MXFP4 & 43.8 & 72.1 & 62.3 & 78.1 & 76.2 & 61.1 & 65.6 \\
        \rowcolor[HTML]{ECF4FF} \textbf{MiX-FP4} & 44.0 & 73.6 & 65.6 & 78.9 & 76.9 & 63.4 & 67.1 \\
        MXINT4 & 42.3 & 68.5 & 62.6 & 76.5 & 74.7 & 58.6 & 63.9 \\
        \rowcolor[HTML]{ECF4FF} \textbf{MiX-INT4} & 46.2 & 73.6 & 64.8 & 79.0 & 77.8 & 64.3 & 67.6 \\
        \midrule
        \multicolumn{8}{c}{\textbf{MiniCPM-V-2.6}} \\
        \midrule
        FP16 & 45.7 & 80.3 & 72.3 & 78.5 & 78.2 & 61.5 & 69.4 \\
        \cmidrule{1-8}
        \multicolumn{8}{c}{\textit{Sparsity-aware / 4.88b}} \\
        Focus & 44.0 & 40.0 & 68.7 & 67.0 & 38.1 & 61.2 & 53.2 \\
        \rowcolor[HTML]{ECF4FF} \textbf{MiX-INT5} & 43.0 & 78.0 & 72.6 & 77.6 & 76.0 & 60.4 & 67.9 \\
        \cmidrule{1-8}
        \multicolumn{8}{c}{\textit{4.50b}} \\
        MXFP4$_{g16}$ & 38.3 & 72.2 & 68.3 & 73.8 & 73.2 & 55.6 & 63.6 \\
        \rowcolor[HTML]{ECF4FF} \textbf{MiX-FP4$_{g16}$} & 40.0 & 75.9 & 70.5 & 75.2 & 74.7 & 57.1 & 65.6 \\
        MXINT4$_{g16}$ & 40.2 & 73.9 & 71.5 & 75.2 & 72.5 & 60.3 & 65.6 \\
        \rowcolor[HTML]{ECF4FF} \textbf{MiX-INT4$_{g16}$} & 41.8 & 74.4 & 72.7 & 76.3 & 73.8 & 61.9 & 66.8 \\
        NVFP4 & 38.7 & 73.0 & 70.6 & 76.9 & 73.4 & 56.9 & 64.9 \\
        MXFP4+ & 42.4 & 77.7 & 69.9 & 76.5 & 77.1 & 60.4 & 67.3 \\
        AMXFP4 & 39.2 & 77.9 & 70.2 & 76.7 & 75.6 & 61.0 & 66.8 \\
        \cmidrule{1-8}
        \multicolumn{8}{c}{\textit{4.25b}} \\
        MXFP4 & 39.0 & 71.5 & 68.4 & 73.3 & 72.3 & 56.3 & 63.5 \\
        \rowcolor[HTML]{ECF4FF} \textbf{MiX-FP4} & 40.2 & 73.6 & 71.3 & 73.6 & 74.9 & 54.9 & 64.7 \\
        MXINT4 & 37.9 & 68.2 & 69.4 & 72.0 & 70.1 & 55.0 & 62.1 \\
        \rowcolor[HTML]{ECF4FF} \textbf{MiX-INT4} & 40.3 & 75.0 & 71.7 & 74.5 & 73.9 & 59.8 & 65.9 \\
        \bottomrule
    \end{tabular}
    }
\vspace{\fpad}
\end{table}

\subsection{Hardware Implementation} 
All proposed and baseline accelerators are implemented at the Register Transfer Level (RTL) in SystemVerilog and synthesized with Synopsys Design Compiler, targeting the TSMC28HPC technology node at a 500\,MHz clock frequency. Power is reported by SAIF annotated simulation with same stimulus to all accelerators.

For end-to-end system evaluation, we deploy a custom cycle-level simulator whose latency and active-energy parameters are derived from the RTL synthesis reports. The memory hierarchy reflects a constrained edge environment: each accelerator is equipped with 288\,KB of activation SRAM, 288\,KB of weight SRAM, and 128\,KB of accumulator memory, with area and read/write energy taken from instance-specific manuals generated by the ARM memory compiler. Off-chip memory is modeled with LPDDR5X specification in \mbox{Ramulator~\cite{ramulator}}.

\begin{table}[!t]
    \centering
    \footnotesize
    \caption{Static prefill-frozen K-mean (Stat) vs.\ dynamic full-sequence K-mean (Dyn, the scheme used in \textcolor{refblue}{Table}~\ref{tab:accuracy_results}).}
    \label{tab:k_smooth_static}
    \begin{tabular}{l cc cc cc}
        \toprule
        & \multicolumn{2}{c}{\textbf{MMMU}} & \multicolumn{2}{c}{\textbf{OCRBench}} & \multicolumn{2}{c}{\textbf{SEED2+}} \\
        \cmidrule(lr){2-3} \cmidrule(lr){4-5} \cmidrule(lr){6-7}
        \textbf{Format} & Dyn & Stat & Dyn & Stat & Dyn & Stat \\
        \midrule
        \multicolumn{7}{c}{\textit{Qwen2-VL-7B}} \\
        MXINT4$_{g16}$ & 43.0 & 42.8 & 74.9 & 74.9 & 63.2 & 63.1 \\
        \rowcolor[HTML]{ECF4FF}
        \textbf{MiX-INT4$_{g16}$} & 46.3 & 44.6 & 78.1 & 76.5 & 65.1 & 66.0 \\
        MXFP4$_{g16}$ & 44.7 & 44.8 & 74.5 & 73.9 & 62.5 & 62.4 \\
        NVFP4 & 45.0 & 45.7 & 78.9 & 76.6 & 65.2 & 66.2 \\
        \midrule
        \multicolumn{7}{c}{\textit{MiniCPM-V-2.6}} \\
        MXINT4$_{g16}$ & 40.2 & 40.2 & 73.9 & 73.0 & 60.3 & 60.1 \\
        \rowcolor[HTML]{ECF4FF}
        \textbf{MiX-INT4$_{g16}$} & 41.8 & 41.3 & 74.4 & 74.5 & 61.9 & 61.5 \\
        MXFP4$_{g16}$ & 38.3 & 39.7 & 72.2 & 73.0 & 55.6 & 56.5 \\
        NVFP4 & 38.7 & 38.0 & 73.0 & 73.5 & 56.9 & 57.0 \\
        \bottomrule
    \end{tabular}
    
\end{table}

\section{Evaluation}
\label{sec:results}

\subsection{End-to-End Multi-Modal Task Accuracy}
\label{sec:eval_accuracy}

To validate the algorithmic fidelity of the proposed co-design, we evaluate six multi-modal benchmarks across three VLMs. Comprehensive results are reported in \textcolor{refblue}{Table}~\ref{tab:accuracy_results}.

\vspace{1mm}\noindent\textbf{The Attention Collapse and K-Smoothing Recovery.}
During initial profiling of Qwen2-VL-7B, we observed catastrophic quantization collapse across all sub-8-bit formats: NVFP4 degraded to an unusable 29.28\% on ChartQA, and MXFP4$_{g16}$ collapsed entirely to 0.0\%. The same collapse on MiniCPM-V-2.6 and LLaVA-OneVision is observed. Deep profiling traced the failure to the Key matrices in the attention mechanism, whose massive channel-wise bias triggers catastrophic microscaling collapse after quantization and destroys the attention map.

To resolve this, we adopt the training-free K-smoothing technique from SageAttention \cite{sageattention}. Because the $K$ outliers manifest as large shared biases across tokens, the matrix can be smoothed by subtracting its token-averaged mean: $\gamma(K) = K - \text{mean}(K)$. Since softmax is invariant to constant shifts, $\sigma(qK^\top - q \cdot \text{mean}(K)) = \sigma(qK^\top)$, leaving the final probability distribution intact. This single transformation fully recovers three models' accuracy and can be fused into the output-channel quantizer with negligible area and latency overhead.

Recomputing the running mean over the full sequence at every decode step would force the accelerator to re-read and re-quantize the entire Key cache as the mean drifts. We propose static K-mean: compute the per-channel mean once at prefill, averaging each Key channel over its \mbox{$N$} prefill tokens, and freeze it in decoding. The flow therefore reduces to an FP32 accumulator that sums each channel, a multiplier by \mbox{$1/N$}, and a single per-channel subtractor that centers each key against the frozen mean. As reported in \mbox{\textcolor{refblue}{Table}~\ref{tab:k_smooth_static}}, this static scheme matches the dynamic full-sequence mean used in \mbox{\textcolor{refblue}{Table}~\ref{tab:accuracy_results}} across both models and all four formats (mean change below 1\,pt, no collapse), and its area and power cost is included in \mbox{\textcolor{refblue}{Table}~\ref{tab:mix_accelerator_overall}}.

\vspace{1mm}\noindent\textbf{End-to-End Quantization vs.\ Prior Art.}
Recent PTQ methods such as MBQ \cite{MBQ} and TLQ \cite{TLQ} quantize only the LLM backbone, leaving the vision encoder and projector in FP16, and reach roughly $60$--$61.5\%$ average accuracy on Qwen2-VL. MiX-INT5 quantizes the \textit{entire} model end-to-end, achieving $70.8\%$ on Qwen2-VL, $67.9\%$ on MiniCPM-V, and $62.4\%$ on LLaVA-OneVision, demonstrating that MiX generalizes across distinct VLM architectures.

\vspace{1mm}\noindent\textbf{The Sub-8-bit Arena.}
Across both Tier 4.5b and Tier 4.25b, every MiX variant improves over its corresponding MX baseline regardless of model or EBW. MiX-INT4$_{g16}$ surpasses MXINT4$_{g16}$ by $2.1\%$ on Qwen2-VL, $1.2\%$ on MiniCPM-V, and $1.6\%$ on LLaVA-OneVision. The same pattern holds for MiX-FP4 over MXFP4 across all tiers. This uniform gain confirms that replacing the shared exponent with per-element exponents universally improves representational fidelity. Against the more expensive NVFP4, MiX-INT4$_{g16}$ matches accuracy on Qwen2-VL ($69.3\%$ vs.\ $69.7\%$) and surpasses it by $1.9\%$ on MiniCPM-V and $0.6\%$ on LLaVA-OneVision, while running on a far simpler shifter-based PE datapath. \mbox{MiX-INT4$_{g16}$} reaches accuracy comparable to the recent outlier-aware FP4 baselines \mbox{MXFP4+~\cite{MX_PLUS}} and \mbox{AMXFP4~\cite{amxfp4}} (with \mbox{$\sim\!1$} pt on the 4.5b tier), but at markedly less hardware cost: AMXFP4's two asymmetric FP8 scales make its PE \mbox{$\sim\!2\times$} larger, and MXFP4+ is likewise less PE-efficient than MiX, so MiX delivers the similar accuracy at higher hardware efficiency.

\vspace{1mm}\noindent\textbf{Focus vs.\ MiX: Sparsity Breaks on Text-Rich VLM Workloads.}
We additionally evaluate \textit{Focus} \cite{focus} on the same six benchmarks under its authors' default sparsity-pruning configuration. On benchmarks dominated by spatial redundancy (SEED-Bench-2+, VizWiz, MMMU\footnote{Focus supports only the single-image subset of MMMU. The accuracy is normalized by the subset size.}), Focus loses at most $3.8$ pts. However, on text-rich workloads where every glyph patch is semantically unique, the same configuration deletes $\sim\!96\%$ of the visual signal, collapsing ChartQA by $21$--$40$ pts and OCRBench by $28$--$40$ pts across LLaVA-OneVision-7B and MiniCPM-V-2.6, dragging MiniCPM-V's six-task average down to $53.2\%$ ($-16.2$ pts vs.\ FP16). MiX, by contrast, dominates Focus on document and chart benchmarks while preserving comparable accuracy on spatial-redundancy workloads.

\begin{table}[!t]
    \centering
    \footnotesize
    \caption{Floor-vs-ceil shared-exponent fix for MXFP4. MiX-FP4 is in shaded background.}
    \label{tab:mxfp4_anomaly}
    \resizebox{\columnwidth}{!}{%
    \begin{tabular}{l c c c c | c c c c}
        \toprule
        & \multicolumn{4}{c|}{\textbf{Tier 4.25b ($B{=}32$)}}
        & \multicolumn{4}{c}{\textbf{Tier 4.5b ($B{=}16$)}} \\
        & \multicolumn{2}{c}{Qwen2-VL} & \multicolumn{2}{c|}{LLaVA-OV}
        & \multicolumn{2}{c}{Qwen2-VL} & \multicolumn{2}{c}{LLaVA-OV} \\
        \cmidrule(lr){2-3} \cmidrule(lr){4-5} \cmidrule(lr){6-7} \cmidrule(lr){8-9}
        \textbf{Format} & CQA & VW & CQA & VW & CQA & VW & CQA & VW \\
        \midrule
        MXINT4         & 74.7 & 62.6 & 70.3 & 58.1 & 78.4 & 63.9 & 74.6 & 58.2 \\
        MXFP4-floor  & 74.2 & 61.5 & 63.8 & 59.5 & 72.9 & 62.1 & 60.4 & 59.2 \\
        \rowcolor[HTML]{ECF4FF}
        MiX-FP4-floor  & 76.3 & 64.4 & 69.0 & 56.6 & 78.6 & 64.8 & 70.3 & 58.7 \\
        MXFP4-ceil     & 76.2 & 62.3 & 70.5 & 61.0 & 76.5 & 63.3 & 71.3 & 61.3 \\
        \rowcolor[HTML]{ECF4FF}
        MiX-FP4-ceil   & 76.9 & 65.6 & 72.0 & 61.3 & 78.9 & 65.4 & 73.8 & 61.4 \\
        \bottomrule
    \end{tabular}}
    
\vspace{\fpad}
\end{table}

\begin{table}[!t]
    \centering
    \footnotesize
    \caption{Area and power breakdown of the proposed MiX-INT4$_{g16}$ 28nm accelerator operating at 500 MHz. A design that the PE with natural block size 16 (16 MACs/PE/cycle) is reported in this table. Area and power are synthesis results.}
    \label{tab:mix_accelerator_overall}
    \begin{tabular}{l c c}
        \toprule
        \textbf{Component} & \textbf{Area ($\mu m^2$)} & \textbf{Power (mW)} \\
        \midrule
        1024-MAC Systolic Array & 130118 & 40.4 \\
        Quantizer & 2540 & 0.795 \\
        K-Smoother & 6258 & 1.8 \\
        SRAM (704KB) & 1715664 & 56.8 \\
        \midrule
        \textbf{Total} & \textbf{1854580} & \textbf{99.795} \\
        \bottomrule
    \end{tabular}
    
\end{table}

\vspace{1mm}\noindent\textbf{The MXFP4 Anomaly.}
Vanilla MXFP4 unexpectedly underperforms MXINT4 across both tiers. We trace this to the OCP spec's \emph{floor}-based shared exponent: with E2M1 elements (max $=6.0$) and an E8M0 power-of-two scale, $\lfloor\log_2(\max|x|)\rfloor$ leaves the post-shared-exp block max in $[4,8)$, but $[6,8)$ is unrepresentable, silently clipping the largest element in roughly $40\%$ of blocks. \textcolor{refblue}{Table}~\ref{tab:mxfp4_anomaly} reports three findings after replacing $\lfloor\cdot\rfloor$ with $\lceil\cdot\rceil$. \textbf{(i) The fix is large}: MXFP4 ChartQA accuracy jumps by up to $+10.9$ pts (LLaVA-OneVision MXFP4$_{g16}$). \textbf{(ii) Block size selects INT vs.\ FP-ceil}: at 4.25b ($B{=}32$), MXFP4-ceil's wider per-element range generally beats MXINT4 (e.g., $+2.9$ pts on LLaVA VizWiz); at 4.5b ($B{=}16$), the smaller block already constrains the intra-block range and uniform-grid MXINT4$_{g16}$ wins instead (e.g., $+3.3$ pts on LLaVA ChartQA). \textbf{(iii) MiX is additive on top of the fix}: MiX-FP4-ceil still beats MXFP4-ceil at every operating point by up to $3.3$ pts, confirming that MiX's per-element exponents capture a representational gain independent of, and additive with, the OCP spec correction.  We consistently report MXFP4-ceil results for MXFP4 format in all tables.

\begin{table}[!t]
    \centering
    \caption{RTL synthesis results (28nm, 500MHz) of the iso-throughput 512-MAC systolic arrays. Power is SAIF-annotated.}
    \label{tab:hardware_comparison}
    \resizebox{\dimexpr0.85\columnwidth-9pt\relax}{!}{
    \begin{tabular}{l c c | c c}
        \toprule
        \textbf{Architecture} & \textbf{Area} & \textbf{Power} & \textbf{Area Eff.} & \textbf{Power Eff.} \\
         & ($\mu m^2$) & (mW) & (TOPS/mm$^2$) & (TOPS/W) \\
        \midrule
        FP16 Baseline & 734236 & 275.8 & 0.7 & 1.9 \\
        INT8 Baseline & 95843 & 61.4 & 5.3 & 8.3 \\
        \midrule
        \multicolumn{5}{c}{\textbf{Tier 4.5b}} \\
        \midrule
        MXINT4$_{g16}$ & 56668 & 34.2 & 9.0 & 15.0 \\
        MXFP4$_{g16}$ & 57422 & 33.9 & 8.9 & 15.1 \\
        MXFP4+ & 58394 & 32.5 & 8.8 & 15.8 \\
        AMXFP4 & 115380 & 47.7 & 4.4 & 10.7 \\
        NVFP4 & 70391 & 36.5 & 7.3 & 14.0 \\
        \rowcolor[HTML]{ECF4FF}
        \textbf{MiX-INT4$_{g16}$} & \textbf{56125} & \textbf{31.8} & \textbf{9.1} & \textbf{16.1} \\
        \rowcolor[HTML]{ECF4FF}
        \textbf{MiX-FP4$_{g16}$} & \textbf{65458} & \textbf{37.1} & \textbf{7.8} & \textbf{13.8} \\
        \midrule
        \multicolumn{5}{c}{\textbf{Tier 4.25b}} \\
        \midrule
        MXINT4 & 47336 & 28.4 & 10.8 & 18.0 \\
        MXFP4 & 50416 & 29.5 & 10.2 & 17.4 \\
        \rowcolor[HTML]{ECF4FF}
        \textbf{MiX-INT4} & \textbf{46407} & \textbf{26.8} & \textbf{11.0} & \textbf{19.1} \\
        \rowcolor[HTML]{ECF4FF}
        \textbf{MiX-FP4} & \textbf{51998} & \textbf{28.5} & \textbf{9.8} & \textbf{18.0} \\
        \bottomrule
    \end{tabular}
    }
\end{table}


\begin{figure*}[!t]
  \centering
  \includegraphics[width=\dimexpr\textwidth-9pt\relax]{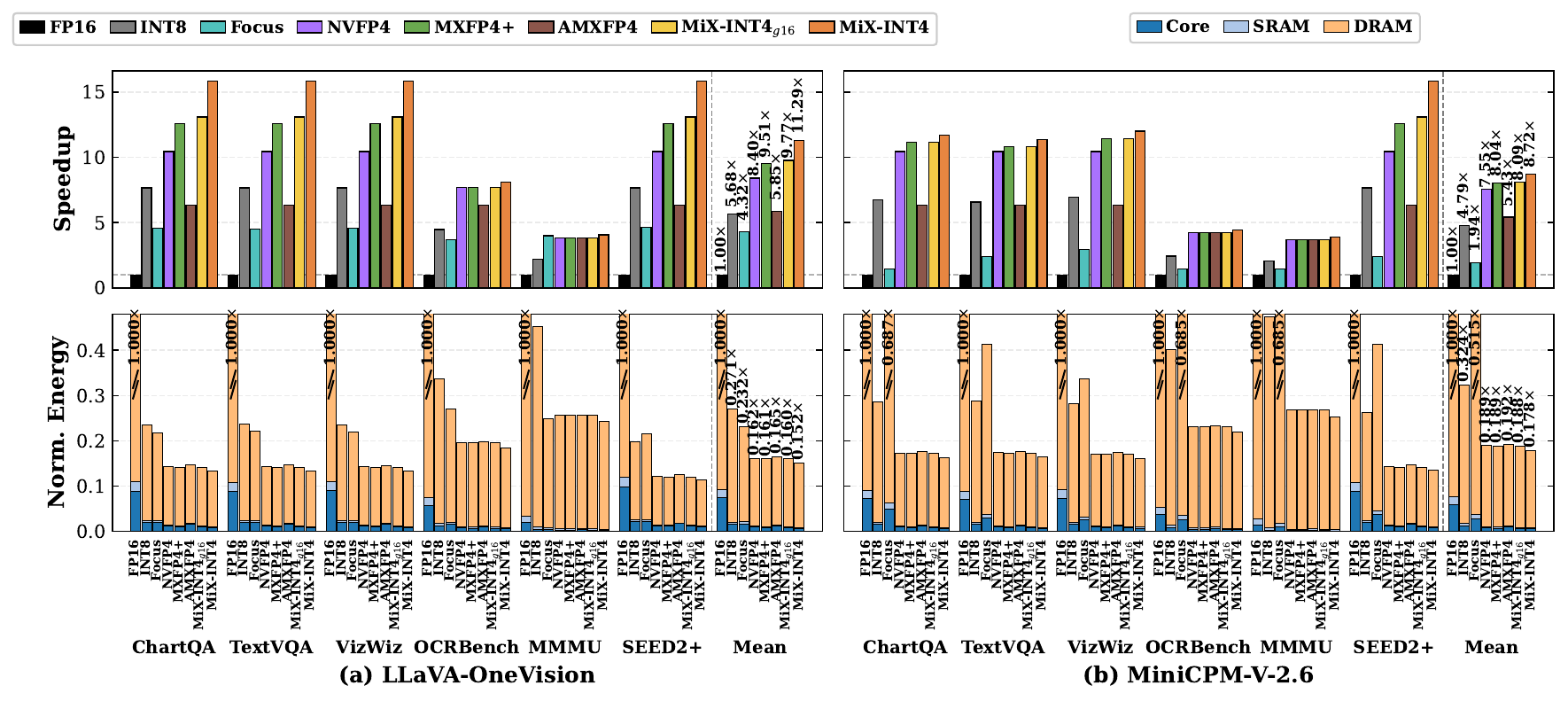}
  \caption{Iso-area speedup and normalized energy breakdown. MMMU is run to its 128-token generation cap (chain-of-thought, memory-bound). The other benchmarks use their typical short generation (1--2 tokens, compute-bound). Mean is the geomean over all six benchmarks.}
  \label{fig:speedup_energy}
  \vspace{+3pt}
\end{figure*}

\subsection{Hardware Efficiency}
\label{sec:eval_hardware}

\textcolor{refblue}{Table}~\ref{tab:mix_accelerator_overall} presents the area and power breakdown of the proposed MiX-INT4$_{g16}$ accelerator. The 1024-MAC systolic array dominates compute at $130118\,\mu m^2$ and $40.4$ mW, while the dual-format quantizer adds only $2540\,\mu m^2$ ($1.9\%$ of the array area) and $0.795$ mW. The K-Smoother is universal to all format and costs $6258\,\mu m^2$ and $1.8$ mW. The $704$\,KB SRAM accounts for the majority of total system area and $56.9\%$ of total system power, consistent with memory-dominated edge accelerator designs.

The complete microarchitectural comparison against baseline formats appears in \textcolor{refblue}{Table}~\ref{tab:hardware_comparison}. The table includes only the compute units, omitting peripheral components such as memory and quantizer, since the compute units dominate area and power while the memory hierarchy is held constant across all configurations.

\vspace{1mm}\noindent\textbf{Tier 4.5b.}
At iso-throughput, \mbox{MiX-INT4$_{g16}$} matches the footprint of \mbox{MXINT4$_{g16}$} (\mbox{$56125$} vs.\ \mbox{$56668\,\mu m^2$}), with near-identical area efficiency (\mbox{$9.1$} vs.\ \mbox{$9.0$} \mbox{TOPS/mm$^2$}) and higher power efficiency (\mbox{$16.1$} vs.\ \mbox{$15.0$} \mbox{TOPS/W}). Against NVFP4, \mbox{MiX-INT4$_{g16}$} is \mbox{$25\%$} more area-efficient and \mbox{$15\%$} more power-efficient (\mbox{$9.1$} vs.\ \mbox{$7.3$} and \mbox{$16.1$} vs.\ \mbox{$14.0$}, respectively) at equal or better task accuracy. The per-element exponent logic therefore adds negligible hardware cost over standard microscaling while substantially closing the accuracy gap to NVFP4's far more expensive two-level scaling architecture. The two outlier aware FP4 baselines are costlier in silicon: AMXFP4 carries two FP8 block scales that roughly double its PE area, dropping it to \mbox{$4.4$} \mbox{TOPS/mm$^2$} and \mbox{$10.7$} \mbox{TOPS/W}, while MXFP4+ adds block maximum extra metadata at \mbox{$8.8$} \mbox{TOPS/mm$^2$} and \mbox{$15.8$} \mbox{TOPS/W}. \mbox{MiX-INT4$_{g16}$} stays more efficient than both on area and power, reaching comparable accuracy at lower hardware cost.

\vspace{1mm}\noindent\textbf{Tier 4.25b.}
At Tier 4.25b, MiX-INT4 achieves the best area efficiency (\mbox{$11.0$ TOPS/mm$^2$}) and the best power efficiency (\mbox{$19.1$ TOPS/W}), simultaneously improving both metrics over the standard MXINT4 baseline while boosting average task accuracy by 3.7 pts on Qwen2-VL. For the MiX-FP4 configuration, a marginal 3.9\% decrease in area efficiency yields a 1.5 pt accuracy gain over MXFP4. Across Tier 4.25b, the MiX overhead is consistently bounded: the per-element exponent shifters add at most 4\% area-efficiency cost relative to the MX baselines, making MiX a drop-in upgrade with substantial accuracy returns.

\subsection{Speedup and Energy}
\label{sec:eval_system}

\mbox{\textcolor{refblue}{Figure}~\ref{fig:speedup_energy}} reports iso-area speedup and normalized energy on LLaVA-OneVision-7B and MiniCPM-V-2.6 across six benchmarks, with MMMU evaluated at its 128-token generation cap to represent long-form chain-of-thought inference. We compare MiX against FP16, INT8, NVFP4, the FP4 outlier-aware baselines MXFP4+ and AMXFP4, and the recent VLM accelerator \textit{Focus}.

\vspace{1mm}\noindent\textbf{The Iso-Area Evaluation Framework.}
A naive cycle-count comparison at identical raw throughput misrepresents architectural efficiency, since Tier 4.5b PEs are \mbox{$6$--$13\times$} smaller than FP16 (\textcolor{refblue}{Section}~\ref{sec:eval_hardware}). We therefore evaluate latency under an \textit{iso-area constraint}: each format is allocated PEs in proportion to its area savings, scaling compute cycles per layer down accordingly, while DRAM cycles remain bounded by the system memory bandwidth. Per-layer latency is the maximum of the scaled compute and DRAM cycles.

\vspace{1mm}\noindent\textbf{End-to-End Speedup: Compute- vs.\ Memory-Bound.}
Short-answer benchmarks generate only 1--2 tokens, so inference is dominated by the vision encoder and prefill and is compute-bound. There, MiX's denser shifter-based PEs convert directly into iso-area throughput, reaching up to \mbox{$13.1\times$} (\mbox{MiX-INT4$_{g16}$}) and \mbox{$15.8\times$} (MiX-INT4) over FP16 on LLaVA-OneVision. When MMMU is run to its 128-token cap (chain-of-thought), the decode phase re-reads all weights every step and the workload turns DRAM-bandwidth-bound, so every 4-bit format converges to the data-movement limit (\mbox{$\sim\!3.8\times$}) regardless of PE size. Averaged across the six benchmarks, \mbox{MiX-INT4$_{g16}$} reaches \mbox{$9.77\times$} on LLaVA-OneVision and \mbox{$8.09\times$} on MiniCPM-V, ahead of NVFP4 (\mbox{$8.40\times$}, \mbox{$7.55\times$}), MXFP4+ (\mbox{$9.51\times$}, \mbox{$8.04\times$}), and AMXFP4 (\mbox{$5.85\times$}, \mbox{$5.43\times$}), whose \mbox{$2\times$} larger dual-scale PE makes it the slowest 4-bit format. On the compute-bound short-generation benchmarks its throughput (\mbox{$6.36\times$}) falls below even INT8 (\mbox{$7.66\times$}). The Tier 4.25b MiX-INT4, with the smallest PE, is highest at \mbox{$11.29\times$} and \mbox{$8.72\times$}.

\vspace{1mm}\noindent\textbf{Comparison with Focus.}
\textit{Focus} achieves speedup by pruning visual tokens before they reach the LLM, but its gains are bounded by how many vision tokens the model produces in the first place. On LLaVA-OneVision (729 vision tokens per image), Focus reaches an average \mbox{$4.32\times$} speedup at \mbox{$0.232\times$} normalized energy. On MiniCPM-V-2.6, whose Q-former resampler already compresses each image slice to only 64 tokens, Focus has far less redundancy to exploit and collapses to a \mbox{$1.94\times$} average speedup at \mbox{$0.515\times$} normalized energy, worse than even the INT8 baseline. By contrast, MiX is a hardware/format attack orthogonal to token count, and its \mbox{$8$--$11\times$} speedup is consistent across both architectures, making MiX a strict win on compact-token VLMs where algorithm-level token pruning is structurally limited.

\vspace{1mm}\noindent\textbf{End-to-End Energy.}
DRAM traffic dominates total energy. By compressing 16-bit operands to \mbox{$\sim$4} bits, every 4-bit format cuts memory energy by \mbox{$5.3\times$--$6.3\times$}. MiX additionally reduces core energy by replacing FP multipliers with shifters. The combined effect places \mbox{MiX-INT4$_{g16}$} at \mbox{$\sim\!0.160\times$} FP16 energy on LLaVA-OneVision and \mbox{$\sim\!0.188\times$} on MiniCPM-V, and \mbox{MiX-INT4} at \mbox{$\sim\!0.152\times$} and \mbox{$\sim\!0.178\times$}, the lowest of any format in either panel, a \mbox{$5.6$--$6.6\times$} end-to-end energy reduction over FP16 and the best energy and area balance in the sub-8-bit arena.

\begin{figure}[!t]
  \centering
  \includegraphics[width=\dimexpr\linewidth-9pt\relax]{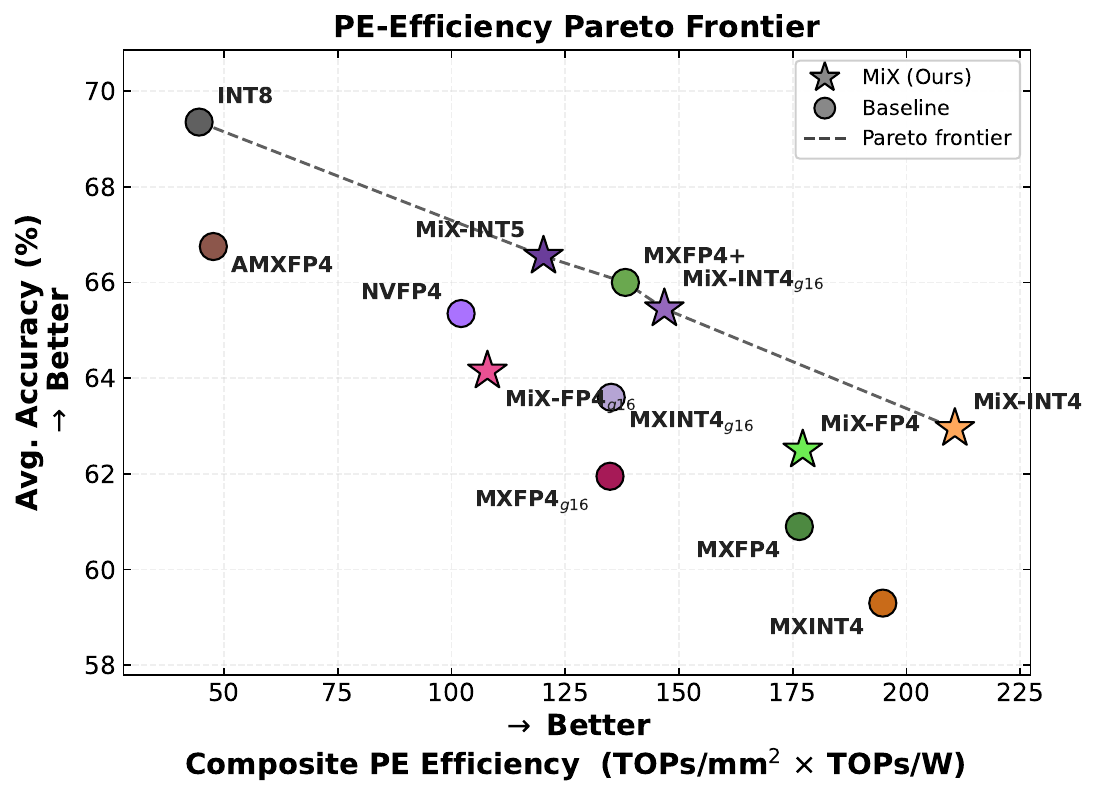}
  \caption{PE-efficiency Pareto frontier across thirteen formats. The dashed line marks the Pareto frontier.}
  \label{fig:pe_efficiency_pareto}
  \vspace{+3pt}
\end{figure}

\subsection{The PPA-Accuracy Pareto Frontier}

\textcolor{refblue}{Figure}~\ref{fig:pe_efficiency_pareto} presents a PE-level Pareto frontier that decouples the format contribution from the memory hierarchy. The $x$-axis is a single composite metric, \mbox{TOPS/mm$^2$ $\times$ TOPS/W}, used for two reasons. First, it rewards a format only when it is both area efficient and power efficient. Second, every format here is either 4.5b or 4.25b, so all share a similar EBW and hence a similar memory footprint and DRAM traffic. Because DRAM access dominates and hides the PE efficiency differences in the energy panel of \mbox{\textcolor{refblue}{Figure}~\ref{fig:speedup_energy}}, this composite metric exposes the intrinsic PE area and power advantage that the bandwidth-bound energy numbers mask. The $y$-axis is the average accuracy across six VLM benchmarks (ChartQA, TextVQA, VizWiz, OCRBench, MMMU, SEED-Bench-2+) and two models (Qwen2-VL-7B and LLaVA-OneVision-7B). We additionally plot INT8 (assumed lossless). \textit{Focus} is not included because of its severe accuracy drop, limited energy reduction, and model-architecture-dependent speedup gain.

\textbf{MiX defines the efficient frontier.} The two most accurate formats, INT8 (\mbox{$69.3\%$}) and AMXFP4 (\mbox{$66.8\%$}), sit at a composite efficiency below \mbox{$50$}, since INT8 uses 8 bits and AMXFP4's two FP8 scales make its PE \mbox{$2\times$} larger, so neither suits efficient sub-8b deployment. In the high-efficiency region (composite above \mbox{$100$}), MiX dominates: \mbox{MiX-INT4$_{g16}$} reaches \mbox{$147$} at \mbox{$65.5\%$}, dominating NVFP4 (\mbox{$102$}, \mbox{$65.3\%$}) on both axes (a \mbox{$44\%$} efficiency gain at higher accuracy), and \mbox{MiX-INT5} lifts accuracy to \mbox{$66.6\%$}. The Tier 4.25b MiX-INT4 extends the frontier rightward to \mbox{$211$} (\mbox{$2.1\times$} NVFP4), the most area and power efficient point in the comparison. The only baseline reaching the frontier alongside MiX is MXFP4+ (\mbox{$138$}, \mbox{$66.0\%$}), which \mbox{MiX-INT4$_{g16}$} trades for higher efficiency at comparable accuracy. Every other format, including NVFP4 and the MX baselines, sits inside it, confirming that per-element exponents improve accuracy and PE efficiency together.

\begin{table}[!t]
    \centering
    \footnotesize
    \caption{Model-size scaling on the Qwen2.5-VL family (3B--72B).}
    \label{tab:vlm_scaling}
    \resizebox{\columnwidth}{!}{%
    \begin{tabular}{l l c c c | c}
        \toprule
        \textbf{Model} & \textbf{Method} & \textbf{MMMU} & \textbf{OCR} & \textbf{SEED2} & \textbf{Avg.} \\
        \midrule
        \multirow{4}{*}{Qwen2.5-VL-3B}
        & FP16 & 46.4 & 80.3 & 67.0 & 64.6 \\
        & MXFP4 & 39.4 & 68.8 & 57.9 & 55.4 \\
        & NVFP4 & 41.2 & 70.3 & 59.4 & 57.0 \\
        \rowcolor[HTML]{ECF4FF}
        \cellcolor{white} & \textbf{MiX-INT4$_{g16}$} & 38.7 & 69.7 & 61.5 & 56.6 \\
        \midrule
        \multirow{4}{*}{Qwen2.5-VL-7B}
        & FP16 & 42.4 & 87.0 & 71.3 & 66.9 \\
        & MXFP4 & 37.1 & 79.0 & 63.5 & 59.9 \\
        & NVFP4 & 38.3 & 81.5 & 68.0 & 62.6 \\
        \rowcolor[HTML]{ECF4FF}
        \cellcolor{white} & \textbf{MiX-INT4$_{g16}$} & 38.6 & 80.8 & 69.1 & 62.8 \\
        \midrule
        \multirow{4}{*}{Qwen2.5-VL-32B}
        & FP16 & 43.3 & 81.1 & 71.9 & 65.4 \\
        & MXFP4 & 40.9 & 71.8 & 67.7 & 60.1 \\
        & NVFP4 & 39.2 & 76.7 & 70.2 & 62.0 \\
        \rowcolor[HTML]{ECF4FF}
        \cellcolor{white} & \textbf{MiX-INT4$_{g16}$} & 41.9 & 76.3 & 70.8 & 63.0 \\
        \midrule
        \multirow{4}{*}{Qwen2.5-VL-72B}
        & FP16 & 61.9 & 86.6 & 73.0 & 73.8 \\
        & MXFP4 & 58.3 & 81.7 & 70.5 & 70.2 \\
        & NVFP4 & 53.7 & 81.9 & 71.6 & 69.1 \\
        \rowcolor[HTML]{ECF4FF}
        \cellcolor{white} & \textbf{MiX-INT4$_{g16}$} & 58.8 & 83.8 & 71.9 & 71.5 \\
        \bottomrule
    \end{tabular}}
  \vspace{+3pt}
\end{table}

\subsection{Scaling and Generalization}
\label{sec:scaling_generalization}

\textbf{Scaling across model size.} \mbox{\textcolor{refblue}{Table}~\ref{tab:vlm_scaling}} sweeps Qwen2.5-VL from 3B to 72B on MMMU, OCRBench, and SEED-Bench-2+, one benchmark per category (reasoning, text-rich, perception). The \mbox{MiX-INT4$_{g16}$} gap to FP16 shrinks monotonically with scale (\mbox{$8.0$}, \mbox{$4.1$}, \mbox{$2.4$}, \mbox{$2.3$} points at 3B, 7B, 32B, 72B), and \mbox{MiX-INT4$_{g16}$} overtakes NVFP4 beyond 3B, leading by \mbox{$0.2$}, \mbox{$1.0$}, and \mbox{$2.4$} points at 7B, 32B, and 72B while beating MXFP4 at every size. MiX thus scales favorably, matching or surpassing NVFP4 at the larger, deployment-relevant sizes on a far cheaper shifter-based PE.

\textbf{Generalizing across modality.} \mbox{\textcolor{refblue}{Table}~\ref{tab:llm_scaling}} applies the same recipe to four text-only LLMs (Llama-3.1-8B, Mistral-7B-v0.3, Qwen2.5-7B/14B) on three reasoning tasks (ARC-Challenge, HellaSwag, WinoGrande) and two perplexity corpora (WikiText, C4), comparing the matched 4.5b \mbox{MiX-INT4$_{g16}$} and NVFP4. MiX is weaker here than on VLMs yet remains close: across the reasoning tasks it trails NVFP4 by \mbox{$1.3$--$1.7$} points on Mistral and Qwen2.5 and is marginally ahead on Llama-3.1-8B. The gap is task dependent, largest on ARC-Challenge (\mbox{$3$--$4$} points) while WinoGrande favors \mbox{MiX-INT4$_{g16}$} (up to \mbox{$3.8$} points on Llama-3.1-8B). This follows from MiX's design: per-element exponents absorb outliers at a small precision cost, aiding range-heavy tasks and penalizing precision-heavy ones. With its efficiency and strong VLM accuracy, MiX remains a sound choice for edge VLM acceleration.

\begin{table}[!t]
    \centering
    \footnotesize
    \caption{Text-only LLM generalization across four models: reasoning accuracy (\%) and perplexity ($\downarrow$, lower is better).}
    \label{tab:llm_scaling}
    \resizebox{\dimexpr\columnwidth-9pt\relax}{!}{
    \begin{tabular}{l l c c c | c c}
        \toprule
        \textbf{Model} & \textbf{Method} & \textbf{ARC-c} & \textbf{HellaS.} & \textbf{WinoG.} & \textbf{Wiki$\downarrow$} & \textbf{C4$\downarrow$} \\
        \midrule
        \multirow{3}{*}{Llama-3.1-8B}
        & FP16 & 53.5 & 79.1 & 73.6 & 6.14 & 8.95 \\
        & NVFP4 & 45.6 & 74.5 & 62.7 & 7.94 & 11.63 \\
        \rowcolor[HTML]{ECF4FF}
        \cellcolor{white} & \textbf{MiX-INT4$_{g16}$} & 46.0 & 71.3 & 66.5 & 8.93 & 12.85 \\
        \midrule
        \multirow{3}{*}{Mistral-7B-v0.3}
        & FP16 & 52.4 & 80.5 & 73.8 & 5.24 & 7.84 \\
        & NVFP4 & 50.5 & 78.9 & 68.7 & 5.75 & 8.47 \\
        \rowcolor[HTML]{ECF4FF}
        \cellcolor{white} & \textbf{MiX-INT4$_{g16}$} & 47.4 & 78.2 & 68.3 & 5.96 & 8.72 \\
        \midrule
        \multirow{3}{*}{Qwen2.5-7B}
        & FP16 & 51.5 & 78.9 & 72.9 & 6.72 & 10.44 \\
        & NVFP4 & 51.3 & 76.9 & 66.0 & 7.67 & 11.51 \\
        \rowcolor[HTML]{ECF4FF}
        \cellcolor{white} & \textbf{MiX-INT4$_{g16}$} & 47.2 & 75.7 & 67.3 & 7.90 & 11.77 \\
        \midrule
        \multirow{3}{*}{Qwen2.5-14B}
        & FP16 & 59.0 & 83.1 & 75.2 & 5.18 & 9.15 \\
        & NVFP4 & 54.9 & 80.3 & 70.6 & 6.25 & 10.05 \\
        \rowcolor[HTML]{ECF4FF}
        \cellcolor{white} & \textbf{MiX-INT4$_{g16}$} & 51.6 & 78.6 & 70.4 & 6.62 & 10.31 \\
        \bottomrule
    \end{tabular}
    }
\end{table}

\section{Related Work}
\label{sec:related_work}

\vspace{1mm}\noindent\textbf{Quantization for LLMs and VLMs.}
LLM post-training quantization targets unimodal models: weight-only schemes (GPTQ \cite{GPTQ}, AWQ \cite{AWQ}, OmniQuant \cite{OmniQuant}, QuaRot \cite{QuaRot}, SpinQuant \cite{spinquant}) push weights below 4 bits, SmoothQuant \cite{smoothquant} and its successors \cite{ZeroQuant, LLMint8} migrate activation outliers for W8A8, and QAT methods \cite{LLM-QAT, EfficientQAT} recover accuracy at calibration cost. VLM-specific work then tackles the cross-modal range gap, where vision tokens, language tokens, and projector activations occupy markedly different dynamic ranges and a single calibration set is rarely sufficient: MBQ \cite{MBQ} balances the vision and text calibration objectives, Q-VLM \cite{Q-VLM} searches cross-layer mixed precision, MQuant \cite{MQuant} adds token-aware scales for the unbalanced visual stream, and TLQ \cite{TLQ}, P4Q \cite{P4Q}, and VEQ \cite{VEQ} target the projector or the vision encoder in isolation. All of these restrict quantization to the linear layers and tune only the most sensitive components, leaving the projector and the attention BMMs in FP16.

\vspace{1mm}\noindent\textbf{Custom-Datatype and Bit-serial Accelerators.}
Several accelerators specialize their PEs for non-uniform data types. ANT \cite{ANT} adapts the per-group integer or floating-point representation to the value distribution, OliVe \cite{OliVe} pairs each outlier with a victim element, M-ANT \cite{M-ANT} extends ANT to mixed-precision attention, and BitMoD \cite{BitMoD} mixes per-group FP4/FP3 types under a unified bit-serial PE. Bit-serial designs (Stripes \cite{Stripes}, Bit-Pragmatic \cite{Pragmatic}, FPRaker \cite{FPRaker}, BitWave \cite{BitWave}) trade latency for bit-level sparsity but still pay the area cost of variable-precision multipliers or wide accumulators, and outlier-aware accelerators such as OLAccel \cite{OLAccel} and AdaptivFloat \cite{AdaptivFloat} likewise retain a multiplier-centric datapath.

\vspace{1mm}\noindent\textbf{Sparsity and Token-Reduction Accelerators.}
A complementary line attacks the activation memory wall through sparsity. SpAtten \cite{SpAtten} cascades token and head pruning, while LAD \cite{LAD}, DOTA \cite{DOTA}, A$^3$ \cite{A3}, and Sanger \cite{Sanger} exploit dynamic attention or KV-cache sparsity. On the vision side, HeatViT \cite{HeatViT} and AdapTiV \cite{AdapTiV} prune or merge visual tokens via attention scores or sign-bit similarity. The work targeting the same VLMs is \textit{Focus} \cite{focus}, which builds a streaming multilevel concentration unit on an FP16 systolic array to remove redundant vision tokens and therefore reduce computation loads. This leaves per-element compute and the FP16 weight footprint unchanged, so their gains collapse on compact-token models such as MiniCPM-V's 64-token resampler. MiX instead attacks the bit-width and the multiplier directly, delivering uniform gains regardless of token count and dominating Focus on throughput and energy (\textcolor{refblue}{Section}~\ref{sec:eval_system}).

\vspace{1mm}\noindent\textbf{Microscaling Formats and Outlier Handling.} The OCP microscaling standard \mbox{\cite{microscaling}} and NVFP4 \mbox{\cite{nvfp4}} share a single exponent per block, which collapses under VLM outliers. Several recent formats extend MX to tolerate outliers. MXFP4+ \mbox{\cite{MX_PLUS}} repurposes the block-maximum exponent field as an extended mantissa, and AMXFP4 \mbox{\cite{amxfp4}} adds two asymmetric FP8 block scales for sign-skewed activations. Both lift accuracy at roughly 4.5b but enlarge the per-element datapath. MicroScopiQ \mbox{\cite{MicroScopiQ}} keeps outliers at higher precision and prunes the least-important weights to reclaim the extra bits, but relies on multi-precision integer PEs and a dedicated network-on-chip to gather the redistributed outliers. MicroMix \mbox{\cite{micromix}} mixes MXFP4, MXFP6, and MXFP8 channels under per-element error thresholds, reaching near-FP16 quality at about 5 effective bits, yet targets the Blackwell FP4 Tensor Cores rather than an edge accelerator and pays a higher average bit-width. MiX takes the orthogonal route: it inverts the format so that per-element exponents absorb outliers while the shared mantissa factors into a multiplier-free shifter PE, reaching comparable accuracy at strictly higher PE efficiency without mixed precision, pruning, or nested scales.

\section{Conclusion}
\label{sec:conclusion}

We presented \textbf{MiX}, an micro-inverted-scaling format that closes the multi-modal dynamic-range gap in Vision-Language Models by assigning each element its own exponent while sharing the mantissa across a small sub-group. Paired with MX weights in a dual-format multiplier-less systolic-array PE, MiX matches or surpasses NVFP4 accuracy on three 7B--8B VLMs while improving PE area efficiency by \mbox{$25\%$} and power efficiency by \mbox{$15\%$} at the same 4.5b effective bit-width. Comparing to the state-of-the-art accelerator \textit{Focus}, MiX delivers up to \mbox{$4.5\times$} speedup and \mbox{$2.9\times$} energy savings in multi-benchmark multi-model evaluation. MiX therefore dominates microscaling baselines, state-of-the-art accelerators, and algorithm-level quantization approaches on the joint PPA--accuracy frontier, establishing MiX as a hardware-friendly path to end-to-end sub-8-bit VLM deployment.

\section*{Acknowledgment}
This work was supported in part by NSF grant 2403723, and Center for the Co-Design of Cognitive Systems (CoCoSys) and ACE Center for Evolvable Computing under JUMP 2.0, an SRC program sponsored by DARPA..

\renewcommand{\appendixname}{Artifact Appendix}
\appendix
\colorlet{refblue}{black}
\sloppy

\subsection{Abstract}
This artifact reproduces every software and hardware result of MiX. It is a single
repository with three self-contained parts: (i)~\texttt{accuracy}: the
MiX / MX / NVFP4 quantization code, per-benchmark YAML configs, and
evaluation scripts that reproduce the VLM and LLM accuracy \textcolor{refblue}{Tables}~\ref{tab:accuracy_results},
\ref{tab:k_smooth_static}, \ref{tab:mxfp4_anomaly}, \ref{tab:vlm_scaling} and
\ref{tab:llm_scaling}; (ii)~\texttt{hardware\_rtl}: SystemVerilog RTL,
self-checking testbenches, and Synopsys Design~Compiler and SAIF-power scripts for
15 systolic-array datapaths together with the MiX/MX quantizer and the K-smoother,
reproducing the area and power in
\textcolor{refblue}{Tables}~\ref{tab:mix_accelerator_overall} and~\ref{tab:hardware_comparison}; and
(iii)~\texttt{hardware\_model}: a standalone analytical energy/area simulator
that reproduces \textcolor{refblue}{Figures}~\ref{fig:speedup_energy} and~\ref{fig:pe_efficiency_pareto}.
Accuracy results are not provided, and the hardware result is summarized in a
reference CSV. Part~1 needs one CUDA GPU sized to the
model (a 48\,GB RTX~A6000 up to 14B, and a 180\,GB B200 at 32B and 72B),
Part~2 an x86-64 Linux host, and Part~3 a CPU.
Parts~1 and~3 use open-source Python (PyTorch for Part~1; NumPy and Matplotlib for
Part~3), while Part~2 requires the proprietary Synopsys Design~Compiler and VCS
with a TSMC 28\,nm standard-cell library.

\subsection{Artifact check-list (meta-information)}
{\small
\begin{itemize}
  \item {\bf Algorithm:} post-training quantization (PTQ) of VLMs and LLMs with the schemes in \texttt{src/quantization} (MiX, MXINT4, MXFP4, NVFP4, AMXFP4, MX+).
  \item {\bf Program:} PyTorch (PTQ and evaluation); SystemVerilog RTL; a Python analytical hardware model.
  \item {\bf Model (HuggingFace):} VLMs LLaVA-OneVision-7B, Qwen2-VL-7B, MiniCPM-V-2.6, Qwen2.5-VL 3B/7B/32B/72B; LLMs Llama-3.1-8B, Mistral-7B-v0.3, Qwen2.5-7B, Qwen2.5-14B.
  \item {\bf Data set:} VLM benchmarks (ChartQA, MMMU, OCRBench, SEED-Bench-2+, TextVQA, VizWiz) and LLM tasks (ARC-Challenge, HellaSwag, WinoGrande, WikiText, C4), downloaded automatically from HuggingFace.
  \item {\bf Run-time environment:} Linux with CUDA 12.x (Part~1); Synopsys EDA (Part~2); any Python~3 (Part~3).
  \item {\bf Hardware:} one GPU sized to the model --- a 48\,GB RTX~A6000 up to 14\,B, a 180\,GB B200 at 32\,B and 72\,B (an 80\,GB A100 is \emph{not} sufficient at 32\,B); a commercial-EDA host; a CPU.
  \item {\bf Metrics:} task accuracy and perplexity; cell area ($\mu$m$^2$), SAIF power (mW), and efficiency (FLOPs/mm$^2$, FLOPs/mW); iso-area speedup and energy.
  \item {\bf Output:} (Part~1) model accuracy for every quantization scheme; (Part~2) area and power of each systolic-array datapath; (Part~3) analytical energy and speedup per format, derived from the Part~2 power numbers.
  \item {\bf Experiments:} YAML-driven accuracy runs; per-design synthesis and SAIF power; analytical simulator scripts.
  \item {\bf Disk space (approx.):} about 50\,GB per 7 to 8\,B model with its datasets; up to about 1\,TB for the full 3B to 72B sweep.
  \item {\bf Workflow preparation time (approx.):} about 1 hour (environment plus first model and dataset download).
  \item {\bf Experiment completion time (approx.):} about one week on 8$\times$A6000 for the main sweep, plus about two days on a B200 for the 32B and 72B models.
  \item {\bf Publicly available?:} Yes.
  \item {\bf Archived (DOI):} \url{https://doi.org/10.5281/zenodo.21530344}
  \item {\bf Code licenses:} MIT (software and RTL).
  \item {\bf Workflow automation framework used?:} GNU Make (synthesis); shell and YAML (evaluation).
\end{itemize}
}

\subsection{Description}

\subsubsection{How to access}
The artifact is archived on Zenodo at
\url{https://doi.org/10.5281/zenodo.21530344}: three top-level directories (one per part), a top-level
\texttt{README.md}, and a \texttt{dependency.md} that lists all hardware,
software, model, and data requirements.

\subsubsection{Hardware dependencies}
Part~1 needs one CUDA GPU. VRAM scales with model size because PTQ inflates each
layer to FP32 before re-casting to FP16. An RTX~A6000 (48\,GB) covers every model
up to 14\,B (measured peak $\approx$33.5\,GB for Qwen2.5-14B); a B200 (180\,GB) is
needed at 32\,B and 72\,B. At 72\,B the FP32 conversion also
needs up to $\sim$500\,GB system RAM. Part~2 needs an x86-64 host running the
Synopsys tools (no GPU/FPGA). Part~3 is CPU-only.

\subsubsection{Software dependencies}
Part~1 uses conda with Python~3.10 and \texttt{pip install -r requirements.txt}
(\texttt{torch}~2.3.0 for CUDA~12.1, \texttt{transformers}~4.57.6, \texttt{timm},
and \texttt{datasets}), all open-source. Part~2 requires the proprietary Synopsys
Design Compiler and VCS, plus a 28\,nm standard-cell \texttt{.db}/\texttt{.v}
library (the paper used TSMC \texttt{tcbn28hpcplus\allowbreak bwp30p140}). Part~3
uses Python with only NumPy and Matplotlib.

\subsubsection{Data sets}
The VLM benchmarks, the LLM tasks, and the calibration data (\texttt{lmms-lab/POPE},
which embeds COCO images, for the VLMs, and Pile for the LLMs) are downloaded
automatically from HuggingFace on the first run. Part~2 ships its own testbenches,
which generate their stimulus internally, so it needs no external data. Part~3
consumes the data generated by Parts~1 and~2, namely the accuracy
\texttt{table2.json} and the area and power CSV.

\subsubsection{Models}
All models are pulled from HuggingFace (exact IDs in \texttt{dependency.md}).
\texttt{meta-llama/Llama-3.1-8B} is gated and MiniCPM-V-2.6 needs remote-code trust.

\subsection{Installation}
A single environment serves the accuracy pipeline (Part~1) and the analytical
model (Part~3), and neither requires a build step:
\begin{verbatim}
conda create -n mix python=3.10 -y
conda activate mix
pip install -r requirements.txt
\end{verbatim}
Model weights and datasets are not installed by hand. They are fetched from
HuggingFace on first use, so set an \texttt{HF\_TOKEN} and accept the Llama license
beforehand. Part~2 needs no installation beyond a licensed Synopsys toolchain:
edit the single \texttt{SITE CONFIGURATION} block at the top of each
\texttt{syn/<Design>/syn.tcl} and of the \texttt{power\_saif/} scripts to point at
your 28\,nm timing library and its Verilog simulation model.

\subsection{Experiment workflow}
\subsubsection{Part~1 (accuracy).} Every entry in \textcolor{refblue}{Tables}~\ref{tab:accuracy_results} to
\ref{tab:llm_scaling} corresponds to one YAML file under \texttt{config/}, whose
directory name selects the model and the quantization scheme. A run loads the
model, applies PTQ with that scheme, calibrates on a small activation set, and
evaluates one benchmark:
\begin{verbatim}
python vlm/qwen2vl.py \
    --config_dir <vlm_config>.yaml
python llm/multiple_choice.py \
    --config_dir <llm_config>.yaml
\end{verbatim}
Each run writes a JSON result under \texttt{accuracy\_result/}, and
\texttt{generate\_table2\_json.py} aggregates these into
\textcolor{refblue}{Table}~\ref{tab:accuracy_results}. The VLM sweep spans six benchmarks (MMMU,
OCRBench, VizWiz, TextVQA, ChartQA, SEED-Bench-2+) and the LLM sweep five tasks.
Calibration draws 128 COCO images for the VLMs and 2{,}048 Pile segments for the
LLMs, and a shell loop over \texttt{config/} reproduces an entire table.

\subsubsection{Part~2 (RTL).} For each of the 15 systolic-array datapaths, \texttt{make
syn} runs Design Compiler for area and timing, and \texttt{power\_saif/run\_all.sh}
obtains SAIF-annotated power from a gate-level simulation with random stimulus;
\texttt{update\_csv.py} collects these into the iso-512 per-format table
(\textcolor{refblue}{Table}~\ref{tab:hardware_comparison}). The accelerator breakdown in
\textcolor{refblue}{Table}~\ref{tab:mix_accelerator_overall} additionally synthesizes the quantizer and
the K-smoother (\texttt{syn/quantizer}, \texttt{syn/k\_smoother}) beside the
systolic array.

\subsubsection{Part~3 (model).} \texttt{mix\_simulator.main} evaluates each workload using
the Part~2 area and power numbers; \texttt{plot\_fig9} produces
\textcolor{refblue}{Figure}~\ref{fig:speedup_energy} and \texttt{pareto.gen\_fig10} produces
\textcolor{refblue}{Figure}~\ref{fig:pe_efficiency_pareto}.

\subsection{Evaluation and expected results}
Part~1 reproduces the reported task accuracy. Representative points are Qwen2-VL-7B
on ChartQA (NVFP4~$=79.2$, MiX-4.5b~$=80.5$) and Qwen2.5-7B under MiX-4.5b
(ARC-Challenge acc\_norm~$=47.2$, WikiText perplexity~$=7.90$). Decoding is greedy,
so results are deterministic within a fixed software environment. At most a
0.1-point difference may appear across library versions on borderline samples.
Further ChartQA reference points are LLaVA-OneVision-7B and MiniCPM-V-2.6 under
NVFP4 (75.6 and 73.4) and Qwen2-VL-7B under AMXFP4 and MiX-4.25b (81.6 and 77.8).
Part~2 reproduces the area and power in
\textcolor{refblue}{Tables}~\ref{tab:mix_accelerator_overall} and~\ref{tab:hardware_comparison}. The
absolute numbers track the standard-cell library, but the ordering across formats,
which is the claim of the paper, is preserved. Every testbench also prints
\texttt{PASS} to confirm that the datapath computes the correct block dot product.
Part~3 regenerates \textcolor{refblue}{Figures}~\ref{fig:speedup_energy}
and~\ref{fig:pe_efficiency_pareto}. Because it derives speedup and energy from the
Part~2 power and the Part~1 accuracy, it closes the loop without a GPU.

\subsection{Experiment customization}
Every accuracy run is driven by its YAML file: the \texttt{wqtype} and
\texttt{xqtype} fields choose the weight and activation schemes, while the block
size, mantissa group, and calibration budget are set beside them, so a new
operating point needs no code change. A genuinely new format is added by
implementing a quantizer in \texttt{src/quantization} and registering it. New
models are registered in \texttt{src/models/auto\_map.py}. On the hardware side
each format is a self-contained directory under \texttt{src/systolic\_arrays} with
a matching \texttt{syn} and \texttt{power\_saif} pair, so an additional datapath is
dropped in and picked up by the same Make flow.

\subsection{Notes}
The three parts are coupled only through generated data:
\textcolor{refblue}{Figure}~\ref{fig:pe_efficiency_pareto} reads the Part~1 accuracy, so build
\texttt{table2.json} first and copy it under \texttt{hardware\_model/} before
running Part~3.

\subsection{Methodology}
Submission, reviewing and badging methodology:
\begin{itemize}
  \item \url{https://www.acm.org/publications/policies/artifact-review-and-badging-current}
  \item \url{https://cTuning.org/ae}
\end{itemize}

\bibliographystyle{IEEEtran}
\bibliography{reference}

@article{MicroScopiQ,
  title={{MicroScopiQ: Accelerating Foundational Models through Outlier-Aware Microscaling Quantization}},
  author={Akshat Ramachandran and Souvik Kundu and Tushar Krishna},
  journal={IEEE/ACM International Symposium on Computer Architecture (ISCA)},
  year={2025}
}

@INPROCEEDINGS{MBQ,
  author={Li, Shiyao and Hu, Yingchun and Ning, Xuefei and Liu, Xihui and Hong, Ke and Jia, Xiaotao and Li, Xiuhong and Yan, Yaqi and Ran, Pei and Dai, Guohao and Yan, Shengen and Yang, Huazhong and Wang, Yu},
  booktitle={IEEE/CVF Conference on Computer Vision and Pattern Recognition (CVPR)}, 
  title={MBQ: Modality-Balanced Quantization for Large Vision-Language Models}, 
  year={2025},
  volume={},
  number={},
  pages={4167-4177},
  doi={10.1109/CVPR52734.2025.00394}}

@inproceedings{AWQ,
 author = {Lin, Ji and Tang, Jiaming and Tang, Haotian and Yang, Shang and Chen, Wei-Ming and Wang, Wei-Chen and Xiao, Guangxuan and Dang, Xingyu and Gan, Chuang and Han, Song},
 booktitle = {Annual Conference on Machine Learning and Systems (MLSys)},
 pages = {87--100},
 title = {AWQ: Activation-aware Weight Quantization for On-Device LLM Compression and Acceleration},
 url = {https://proceedings.mlsys.org/paper_files/paper/2024/file/42a452cbafa9dd64e9ba4aa95cc1ef21-Paper-Conference.pdf},
 year = {2024}
}

@inproceedings{smoothquant,
author = {Xiao, Guangxuan and Lin, Ji and Seznec, Mickael and Wu, Hao and Demouth, Julien and Han, Song},
title = {SmoothQuant: Accurate and efficient post-training quantization for large language models},
year = {2023},
booktitle = {International Conference on Machine Learning (ICML)},
}

@INPROCEEDINGS{focus,
  author={Wei, Chiyue and Guo, Cong and Zhang, Junyao and Shan, Haoxuan and Xu, Yifan and Zhang, Ziyue and Liu, Yudong and Wang, Qinsi and Zhou, Changchun and Li, Hai Helen and Chen, Yiran},
  booktitle={IEEE International Symposium on High Performance Computer Architecture (HPCA)}, 
  title={Focus: A Streaming Concentration Architecture for Efficient Vision-Language Models}, 
  year={2026},
  volume={},
  number={},
  pages={1-18},
  doi={10.1109/HPCA68181.2026.11408525}}

@inproceedings{Q-VLM,
author = {Wang, Changyuan and Wang, Ziwei and Xu, Xiuwei and Tang, Yansong and Zhou, Jie and Lu, Jiwen},
title = {Q-VLM: post-training quantization for large vision-language models},
year = {2024},
isbn = {9798331314385},
booktitle = {Annual Conference on Neural Information Processing Systems (NeurIPS)},
}

@inproceedings{MQuant,
author = {Yu, Jiangyong and Zhou, Sifan and Yang, Dawei and Li, Shuoyu and Wang, Shuo and Hu, Xing and Xu, Chen and Xu, Zukang and Shu, Changyong and Yuan, Zhihang},
title = {MQuant: Unleashing the Inference Potential of Multimodal Large Language Models via Static Quantization},
year = {2025},
isbn = {9798400720352},
url = {https://doi.org/10.1145/3746027.3755433},
doi = {10.1145/3746027.3755433},
booktitle = {ACM International Conference on Multimedia},
pages = {1783–1792},
}

@misc{VEQ,
      title={VEQ: Modality-Adaptive Quantization for MoE Vision-Language Models}, 
      author={Guangshuo Qin and Zhiteng Li and Zheng Chen and Weihang Zhang and Linghe Kong and Yulun Zhang},
      year={2026},
      eprint={2602.01037},
      archivePrefix={arXiv},
      primaryClass={cs.CV},
      url={https://arxiv.org/abs/2602.01037}, 
}

@misc{TLQ,
      title={Rethinking Practical and Efficient Quantization Calibration for Vision-Language Models}, 
      author={Zhenhao Shang and Haizhao Jing and Guoting Wei and Haokui Zhang and Rong Xiao and Jianqing Gao and Peng Wang},
      year={2026},
      eprint={2602.07899},
      archivePrefix={arXiv},
      primaryClass={cs.CV},
      url={https://arxiv.org/abs/2602.07899}, 
}

@inproceedings{spinquant,
      title={SpinQuant: LLM quantization with learned rotations}, 
      author={Zechun Liu and Changsheng Zhao and Igor Fedorov and Bilge Soran and Dhruv Choudhary and Raghuraman Krishnamoorthi and Vikas Chandra and Yuandong Tian and Tijmen Blankevoort},
      year={2025},
      booktitle={International Conference on Learning Representations (ICLR)}
}

@INPROCEEDINGS{BitMoD,
  author={Chen, Yuzong and AbouElhamayed, Ahmed F. and Dai, Xilai and Wang, Yang and Andronic, Marta and Constantinides, George A. and Abdelfattah, Mohamed S.},
  booktitle={IEEE International Symposium on High Performance Computer Architecture (HPCA)}, 
  title={BitMoD: Bit-serial Mixture-of-Datatype LLM Acceleration}, 
  year={2025},
  volume={},
  number={},
  pages={1082-1097},
  doi={10.1109/HPCA61900.2025.00084}}

@INPROCEEDINGS{M-ANT,
  author={Hu, Weiming and Zhang, Haoyan and Guo, Cong and Feng, Yu and Guan, Renyang and Hua, Zhendong and Liu, Zihan and Guan, Yue and Guo, Minyi and Leng, Jingwen},
  booktitle={2025 IEEE International Symposium on High Performance Computer Architecture (HPCA)}, 
  title={M-ANT: Efficient Low-bit Group Quantization for LLMs via Mathematically Adaptive Numerical Type}, 
  year={2025},
  volume={},
  number={},
  pages={1112-1126},
  doi={10.1109/HPCA61900.2025.00086}}

@manual{microscaling,
  title        = {OCP Microscaling Formats (MX) Specification},
  author       = {{Open Compute Project}},
  organization = {Open Compute Project},
  edition      = {Version 1.0},
  year         = {2023},
  month        = {sep},
  url          = {https://www.opencompute.org/documents/ocp-microscaling-formats-mx-v1-0-spec-final-pdf},
  note         = {Accessed: 2026-04-05}
}

@misc{qwen2vl,
      title={Qwen2-VL: Enhancing Vision-Language Model's Perception of the World at Any Resolution}, 
      author={Peng Wang and Shuai Bai and Sinan Tan and Shijie Wang and Zhihao Fan and Jinze Bai and Keqin Chen and Xuejing Liu and Jialin Wang and Wenbin Ge and Yang Fan and Kai Dang and Mengfei Du and Xuancheng Ren and Rui Men and Dayiheng Liu and Chang Zhou and Jingren Zhou and Junyang Lin},
      year={2024},
      eprint={2409.12191},
      archivePrefix={arXiv},
      primaryClass={cs.CV},
      url={https://arxiv.org/abs/2409.12191}, 
}

@misc{llava-onevision,
      title={LLaVA-OneVision: Easy Visual Task Transfer}, 
      author={Bo Li and Yuanhan Zhang and Dong Guo and Renrui Zhang and Feng Li and Hao Zhang and Kaichen Zhang and Peiyuan Zhang and Yanwei Li and Ziwei Liu and Chunyuan Li},
      year={2024},
      eprint={2408.03326},
      archivePrefix={arXiv},
      primaryClass={cs.CV},
      url={https://arxiv.org/abs/2408.03326}, 
}

@misc{minicpm-v,
      title={MiniCPM-V: A GPT-4V Level MLLM on Your Phone}, 
      author={Yuan Yao and Tianyu Yu and Ao Zhang and Chongyi Wang and Junbo Cui and Hongji Zhu and Tianchi Cai and Haoyu Li and Weilin Zhao and Zhihui He and Qianyu Chen and Huarong Zhou and Zhensheng Zou and Haoye Zhang and Shengding Hu and Zhi Zheng and Jie Zhou and Jie Cai and Xu Han and Guoyang Zeng and Dahai Li and Zhiyuan Liu and Maosong Sun},
      year={2024},
      eprint={2408.01800},
      archivePrefix={arXiv},
      primaryClass={cs.CV},
      url={https://arxiv.org/abs/2408.01800}, 
}

@article{OCRBench,
   title={OCRBench: on the hidden mystery of OCR in large multimodal models},
   volume={67},
   ISSN={1869-1919},
   url={http://dx.doi.org/10.1007/s11432-024-4235-6},
   DOI={10.1007/s11432-024-4235-6},
   number={12},
   journal={Science China Information Sciences},
   publisher={Springer Science and Business Media LLC},
   author={Liu, Yuliang and Li, Zhang and Huang, Mingxin and Yang, Biao and Yu, Wenwen and Li, Chunyuan and Yin, Xu-Cheng and Liu, Cheng-Lin and Jin, Lianwen and Bai, Xiang},
   year={2024},
   month=dec }

@inproceedings{MMMU,
      title={MMMU: A Massive Multi-discipline Multimodal Understanding and Reasoning Benchmark for Expert AGI},
      author={Xiang Yue and Yuansheng Ni and Kai Zhang and Tianyu Zheng and Ruoqi Liu and Ge Zhang and Samuel Stevens and Dongfu Jiang and Weiming Ren and Yuxuan Sun and Cong Wei and Botao Yu and Ruibin Yuan and Renliang Sun and Ming Yin and Boyuan Zheng and Zhenzhu Yang and Yibo Liu and Wenhao Huang and Huan Sun and Yu Su and Wenhu Chen},
      booktitle={IEEE/CVF Conference on Computer Vision and Pattern Recognition (CVPR)},
      year={2024},
    }

@misc{VizWiz,
      title={VizWiz Grand Challenge: Answering Visual Questions from Blind People}, 
      author={Danna Gurari and Qing Li and Abigale J. Stangl and Anhong Guo and Chi Lin and Kristen Grauman and Jiebo Luo and Jeffrey P. Bigham},
      year={2018},
      eprint={1802.08218},
      archivePrefix={arXiv},
      primaryClass={cs.CV},
      url={https://arxiv.org/abs/1802.08218}, 
}

@misc{TextVQA,
      title={Towards VQA Models That Can Read}, 
      author={Amanpreet Singh and Vivek Natarajan and Meet Shah and Yu Jiang and Xinlei Chen and Dhruv Batra and Devi Parikh and Marcus Rohrbach},
      year={2019},
      eprint={1904.08920},
      archivePrefix={arXiv},
      primaryClass={cs.CL},
      url={https://arxiv.org/abs/1904.08920}, 
}

@misc{ChartQA,
      title={ChartQA: A Benchmark for Question Answering about Charts with Visual and Logical Reasoning}, 
      author={Ahmed Masry and Do Xuan Long and Jia Qing Tan and Shafiq Joty and Enamul Hoque},
      year={2022},
      eprint={2203.10244},
      archivePrefix={arXiv},
      primaryClass={cs.CL},
      url={https://arxiv.org/abs/2203.10244}, 
}

@misc{SEEDBench2Plus,
      title={SEED-Bench-2-Plus: Benchmarking Multimodal Large Language Models with Text-Rich Visual Comprehension}, 
      author={Bohao Li and Yuying Ge and Yi Chen and Yixiao Ge and Ruimao Zhang and Ying Shan},
      year={2024},
      eprint={2404.16790},
      archivePrefix={arXiv},
      primaryClass={cs.CV},
      url={https://arxiv.org/abs/2404.16790}, 
}

@inproceedings{sageattention,
  title={SageAttention: Accurate 8-Bit Attention for Plug-and-play Inference Acceleration}, 
  author={Zhang, Jintao and Wei, Jia and Zhang, Pengle and Zhu, Jun and Chen, Jianfei},
  booktitle={International Conference on Learning Representations (ICLR)},
  year={2025}
}

@misc{nvfp4,
  author = {{NVIDIA}},
  title = {{Introducing NVFP4 for Efficient and Accurate Low-Precision Inference}},
  url = {https://developer.nvidia.com/blog/introducing-nvfp4-for-efficient-and-accurate-low-precision-inference/}
}

@inproceedings{LLaVA,
 author = {Liu, Haotian and Li, Chunyuan and Wu, Qingyang and Lee, Yong Jae},
 booktitle = {Annual Conference on Neural Information Processing Systems (NeurIPS)},
 pages = {34892--34916},
 title = {Visual Instruction Tuning},
 url = {https://proceedings.neurips.cc/paper_files/paper/2023/file/6dcf277ea32ce3288914faf369fe6de0-Paper-Conference.pdf},
 year = {2023}
}

@InProceedings{InternVL,
    author    = {Chen, Zhe and Wu, Jiannan and Wang, Wenhai and Su, Weijie and Chen, Guo and Xing, Sen and Zhong, Muyan and Zhang, Qinglong and Zhu, Xizhou and Lu, Lewei and Li, Bin and Luo, Ping and Lu, Tong and Qiao, Yu and Dai, Jifeng},
    title     = {InternVL: Scaling up Vision Foundation Models and Aligning for Generic Visual-Linguistic Tasks},
    booktitle = {IEEE/CVF Conference on Computer Vision and Pattern Recognition (CVPR)},
    month     = {June},
    year      = {2024},
    pages     = {24185-24198}
}

@INPROCEEDINGS {SpAtten,
author = { Wang, Hanrui and Zhang, Zhekai and Han, Song },
booktitle = {IEEE International Symposium on High-Performance Computer Architecture (HPCA) },
title = {{ SpAtten: Efficient Sparse Attention Architecture with Cascade Token and Head Pruning}},
year = {2021},
volume = {},
ISSN = {},
pages = {97-110},
doi = {10.1109/HPCA51647.2021.00018},
month =mar}

@INPROCEEDINGS{HeatViT,
  author={Dong, Peiyan and Sun, Mengshu and Lu, Alec and Xie, Yanyue and Liu, Kenneth and Kong, Zhenglun and Meng, Xin and Li, Zhengang and Lin, Xue and Fang, Zhenman and Wang, Yanzhi},
  booktitle={IEEE International Symposium on High-Performance Computer Architecture (HPCA)}, 
  title={HeatViT: Hardware-Efficient Adaptive Token Pruning for Vision Transformers}, 
  year={2023},
  volume={},
  number={},
  pages={442-455},
  doi={10.1109/HPCA56546.2023.10071047}}

@INPROCEEDINGS{AdapTiV,
  author={Yoo, Seungjae and Kim, Hangyeol and Kim, Joo-Young},
  booktitle={IEEE/ACM International Symposium on Microarchitecture (MICRO)}, 
  title={AdapTiV: Sign-Similarity Based Image-Adaptive Token Merging for Vision Transformer Acceleration}, 
  year={2024},
  volume={},
  number={},
  pages={64-77},
  doi={10.1109/MICRO61859.2024.00015}}

@inproceedings{PagedAttention,
author = {Kwon, Woosuk and Li, Zhuohan and Zhuang, Siyuan and Sheng, Ying and Zheng, Lianmin and Yu, Cody Hao and Gonzalez, Joseph and Zhang, Hao and Stoica, Ion},
title = {Efficient Memory Management for Large Language Model Serving with PagedAttention},
year = {2023},
isbn = {9798400702297},
url = {https://doi.org/10.1145/3600006.3613165},
doi = {10.1145/3600006.3613165},
booktitle = {ACM Symposium on Operating Systems Principles},
pages = {611–626},
}

@inproceedings{H2O,
 author = {Zhang, Zhenyu and Sheng, Ying and Zhou, Tianyi and Chen, Tianlong and Zheng, Lianmin and Cai, Ruisi and Song, Zhao and Tian, Yuandong and R\'{e}, Christopher and Barrett, Clark and Wang, Zhangyang "Atlas" and Chen, Beidi},
 booktitle = {Annual Conference on Neural Information Processing Systems (NeurIPS)},
 pages = {34661--34710},
 title = {H2O: Heavy-Hitter Oracle for Efficient Generative Inference of Large Language Models},
 url = {https://proceedings.neurips.cc/paper_files/paper/2023/file/6ceefa7b15572587b78ecfcebb2827f8-Paper-Conference.pdf},
 year = {2023}
}

@misc{GPTQ,
      title={GPTQ: Accurate Post-Training Quantization for Generative Pre-trained Transformers}, 
      author={Elias Frantar and Saleh Ashkboos and Torsten Hoefler and Dan Alistarh},
      year={2023},
      eprint={2210.17323},
      archivePrefix={arXiv},
      primaryClass={cs.LG},
      url={https://arxiv.org/abs/2210.17323}, 
}

@inproceedings{OmniQuant,
title={OmniQuant: Omnidirectionally Calibrated Quantization for Large Language Models},
author={Wenqi Shao and Mengzhao Chen and Zhaoyang Zhang and Peng Xu and Lirui Zhao and Zhiqian Li and Kaipeng Zhang and Peng Gao and Yu Qiao and Ping Luo},
booktitle={International Conference on Learning Representations (ICLR)},
year={2024},
}

@inproceedings{KIVI,
author = {Liu, Zirui and Yuan, Jiayi and Jin, Hongye and Zhong, Shaochen (Henry) and Xu, Zhaozhuo and Braverman, Vladimir and Chen, Beidi and Hu, Xia},
title = {KIVI: a tuning-free asymmetric 2bit quantization for KV cache},
year = {2024},
booktitle = {International Conference on Machine Learning (ICML)},
}

@inproceedings{KVQuant,
 author = {Hooper, Coleman and Kim, Sehoon and Mohammadzadeh, Hiva and Mahoney, Michael W. and Shao, Yakun Sophia and Keutzer, Kurt and Gholami, Amir},
 booktitle = {Annual Conference on Neural Information Processing Systems (NeurIPS)},
 doi = {10.52202/079017-0040},
 pages = {1270--1303},
 title = {KVQuant: Towards 10 Million Context Length LLM Inference with KV Cache Quantization},
 url = {https://proceedings.neurips.cc/paper_files/paper/2024/file/028fcbcf85435d39a40c4d61b42c99a4-Paper-Conference.pdf},
 year = {2024}
}

@inproceedings{LLMint8,
author = {Dettmers, Tim and Lewis, Mike and Belkada, Younes and Zettlemoyer, Luke},
title = {LLM.int8(): 8-bit matrix multiplication for transformers at scale},
year = {2022},
isbn = {9781713871088},
booktitle = {Annual Conference on Neural Information Processing Systems (NeurIPS)},
}

@inproceedings{QuaRot,
 author = {Ashkboos, Saleh and Mohtashami, Amirkeivan and Croci, Maximilian L. and Li, Bo and Cameron, Pashmina and Jaggi, Martin and Alistarh, Dan and Hoefler, Torsten and Hensman, James},
 booktitle = {Annual Conference on Neural Information Processing Systems (NeurIPS)},
 doi = {10.52202/079017-3180},
 pages = {100213--100240},
 title = {QuaRot: Outlier-Free 4-Bit Inference in Rotated LLMs},
 url = {https://proceedings.neurips.cc/paper_files/paper/2024/file/b5b939436789f76f08b9d0da5e81af7c-Paper-Conference.pdf},
 volume = {37},
 year = {2024}
}

@misc{P4Q,
      title={P4Q: Learning to Prompt for Quantization in Visual-language Models}, 
      author={Huixin Sun and Runqi Wang and Yanjing Li and Xianbin Cao and Xiaolong Jiang and Yao Hu and Baochang Zhang},
      year={2024},
      eprint={2409.17634},
      archivePrefix={arXiv},
      primaryClass={cs.CV},
      url={https://arxiv.org/abs/2409.17634}, 
}

@inproceedings{OliVe,
author = {Guo, Cong and Tang, Jiaming and Hu, Weiming and Leng, Jingwen and Zhang, Chen and Yang, Fan and Liu, Yunxin and Guo, Minyi and Zhu, Yuhao},
title = {OliVe: Accelerating Large Language Models via Hardware-friendly Outlier-Victim Pair Quantization},
year = {2023},
isbn = {9798400700958},
url = {https://doi.org/10.1145/3579371.3589038},
doi = {10.1145/3579371.3589038},
booktitle = {IEEE/ACM International Symposium on Computer Architecture (ISCA)},
}

@INPROCEEDINGS{ANT,
  author={Guo, Cong and Zhang, Chen and Leng, Jingwen and Liu, Zihan and Yang, Fan and Liu, Yunxin and Guo, Minyi and Zhu, Yuhao},
  booktitle={IEEE/ACM International Symposium on Microarchitecture (MICRO)}, 
  title={ANT: Exploiting Adaptive Numerical Data Type for Low-bit Deep Neural Network Quantization}, 
  year={2022},
  volume={},
  number={},
  pages={1414-1433},
  doi={10.1109/MICRO56248.2022.00095}}

@inproceedings{Flexpoint,
 author = {K\"{o}ster, Urs and Webb, Tristan and Wang, Xin and Nassar, Marcel and Bansal, Arjun K and Constable, William and Elibol, Oguz and Gray, Scott and Hall, Stewart and Hornof, Luke and Khosrowshahi, Amir and Kloss, Carey and Pai, Ruby J and Rao, Naveen},
 booktitle = {Annual Conference on Neural Information Processing Systems (NeurIPS)},
 title = {Flexpoint: An Adaptive Numerical Format for Efficient Training of Deep Neural Networks},
 url = {https://proceedings.neurips.cc/paper_files/paper/2017/file/a0160709701140704575d499c997b6ca-Paper.pdf},
 year = {2017}
}

@inproceedings{HBFP,
 author = {Drumond, Mario and LIN, Tao and Jaggi, Martin and Falsafi, Babak},
 booktitle = {Annual Conference on Neural Information Processing Systems (NeurIPS)},
 editor = {S. Bengio and H. Wallach and H. Larochelle and K. Grauman and N. Cesa-Bianchi and R. Garnett},
 pages = {},
 title = {Training DNNs with Hybrid Block Floating Point},
 url = {https://proceedings.neurips.cc/paper_files/paper/2018/file/6a9aeddfc689c1d0e3b9ccc3ab651bc5-Paper.pdf},
 year = {2018}
}

@inproceedings{attention,
author = {Vaswani, Ashish and Shazeer, Noam and Parmar, Niki and Uszkoreit, Jakob and Jones, Llion and Gomez, Aidan N. and Kaiser, \L{}ukasz and Polosukhin, Illia},
title = {Attention is all you need},
year = {2017},
isbn = {9781510860964},
booktitle = {Annual Conference on Neural Information Processing Systems (NeurIPS)},
pages = {6000–6010},
numpages = {11},
}

@inproceedings{ZeroQuant,
 author = {Yao, Zhewei and Yazdani Aminabadi, Reza and Zhang, Minjia and Wu, Xiaoxia and Li, Conglong and He, Yuxiong},
 booktitle = {Annual Conference on Neural Information Processing Systems (NeurIPS)},
 pages = {27168--27183},
 title = {ZeroQuant: Efficient and Affordable Post-Training Quantization for Large-Scale Transformers},
 url = {https://proceedings.neurips.cc/paper_files/paper/2022/file/adf7fa39d65e2983d724ff7da57f00ac-Paper-Conference.pdf},
 year = {2022}
}

@misc{LLM-QAT,
      title={LLM-QAT: Data-Free Quantization Aware Training for Large Language Models}, 
      author={Zechun Liu and Barlas Oguz and Changsheng Zhao and Ernie Chang and Pierre Stock and Yashar Mehdad and Yangyang Shi and Raghuraman Krishnamoorthi and Vikas Chandra},
      year={2023},
      eprint={2305.17888},
      archivePrefix={arXiv},
      primaryClass={cs.CL},
      url={https://arxiv.org/abs/2305.17888}, 
}

@misc{EfficientQAT,
      title={EfficientQAT: Efficient Quantization-Aware Training for Large Language Models}, 
      author={Mengzhao Chen and Wenqi Shao and Peng Xu and Jiahao Wang and Peng Gao and Kaipeng Zhang and Ping Luo},
      year={2025},
      eprint={2407.11062},
      archivePrefix={arXiv},
      primaryClass={cs.LG},
      url={https://arxiv.org/abs/2407.11062}, 
}

@inproceedings{Pragmatic,
author = {Albericio, Jorge and Delm\'{a}s, Alberto and Judd, Patrick and Sharify, Sayeh and O'Leary, Gerard and Genov, Roman and Moshovos, Andreas},
title = {Bit-pragmatic deep neural network computing},
year = {2017},
isbn = {9781450349529},
publisher = {Association for Computing Machinery},
address = {New York, NY, USA},
url = {https://doi.org/10.1145/3123939.3123982},
doi = {10.1145/3123939.3123982},
booktitle = {Proceedings of the 50th Annual IEEE/ACM International Symposium on Microarchitecture},
pages = {382–394},
numpages = {13},
location = {Cambridge, Massachusetts},
series = {MICRO-50 '17}
}

@INPROCEEDINGS{Stripes,
  author={Judd, Patrick and Albericio, Jorge and Hetherington, Tayler and Aamodt, Tor M. and Moshovos, Andreas},
  booktitle={2016 49th Annual IEEE/ACM International Symposium on Microarchitecture (MICRO)}, 
  title={Stripes: Bit-serial deep neural network computing}, 
  year={2016},
  volume={},
  number={},
  pages={1-12},
  doi={10.1109/MICRO.2016.7783722}}

@inproceedings{FPRaker,
author = {Awad, Omar Mohamed and Mahmoud, Mostafa and Edo, Isak and Zadeh, Ali Hadi and Bannon, Ciaran and Jayarajan, Anand and Pekhimenko, Gennady and Moshovos, Andreas},
title = {FPRaker: A Processing Element For Accelerating Neural Network Training},
year = {2021},
isbn = {9781450385572},
publisher = {Association for Computing Machinery},
address = {New York, NY, USA},
url = {https://doi.org/10.1145/3466752.3480106},
doi = {10.1145/3466752.3480106},
booktitle = {MICRO-54: 54th Annual IEEE/ACM International Symposium on Microarchitecture},
pages = {857–869},
numpages = {13},
location = {Virtual Event, Greece},
series = {MICRO '21}
}

@inproceedings{BitWave,
   title={BitWave: Exploiting Column-Based Bit-Level Sparsity for Deep Learning Acceleration},
   url={http://dx.doi.org/10.1109/HPCA57654.2024.00062},
   DOI={10.1109/hpca57654.2024.00062},
   booktitle={IEEE International Symposium on High-Performance Computer Architecture (HPCA)},
   author={Shi, Man and Jain, Vikram and Joseph, Antony and Meijer, Maurice and Verhelst, Marian},
   year={2024},
   month=mar, pages={732–746} }

@INPROCEEDINGS{AdaptivFloat,
  author={Tambe, Thierry and Yang, En-Yu and Wan, Zishen and Deng, Yuntian and Janapa Reddi, Vijay and Rush, Alexander and Brooks, David and Wei, Gu-Yeon},
  booktitle={ACM/IEEE Design Automation Conference (DAC)}, 
  title={Algorithm-Hardware Co-Design of Adaptive Floating-Point Encodings for Resilient Deep Learning Inference}, 
  year={2020},
  volume={},
  number={},
  pages={1-6},
  doi={10.1109/DAC18072.2020.9218516}}

@inproceedings{OLAccel,
author = {Park, Eunhyeok and Kim, Dongyoung and Yoo, Sungjoo},
title = {Energy-efficient neural network accelerator based on outlier-aware low-precision computation},
year = {2018},
isbn = {9781538659847},
url = {https://doi.org/10.1109/ISCA.2018.00063},
doi = {10.1109/ISCA.2018.00063},
booktitle = {IEEE/ACM International Symposium on Computer Architecture (ISCA)},
pages = {688–698},
}

@inproceedings{DOTA,
author = {Qu, Zheng and Liu, Liu and Tu, Fengbin and Chen, Zhaodong and Ding, Yufei and Xie, Yuan},
title = {DOTA: detect and omit weak attentions for scalable transformer acceleration},
year = {2022},
isbn = {9781450392051},
url = {https://doi.org/10.1145/3503222.3507738},
doi = {10.1145/3503222.3507738},
booktitle = {ACM International Conference on Architectural Support for Programming Languages and Operating Systems (ASPLOS)},
pages = {14–26},
}

@inproceedings{Sanger,
author = {Lu, Liqiang and Jin, Yicheng and Bi, Hangrui and Luo, Zizhang and Li, Peng and Wang, Tao and Liang, Yun},
title = {Sanger: A Co-Design Framework for Enabling Sparse Attention using Reconfigurable Architecture},
year = {2021},
isbn = {9781450385572},
url = {https://doi.org/10.1145/3466752.3480125},
doi = {10.1145/3466752.3480125},
booktitle = {IEEE/ACM International Symposium on Microarchitecture (MICRO)},
pages = {977-991},
}

@INPROCEEDINGS{LAD,
  author={Wang, Haoran and Li, Yuming and Xu, Haobo and Wang, Ying and Liu, Liqi and Yang, Jun and Han, Yinhe},
  booktitle={IEEE International Symposium on High Performance Computer Architecture (HPCA)}, 
  title={LAD: Efficient Accelerator for Generative Inference of LLM with Locality Aware Decoding}, 
  year={2025},
  volume={},
  number={},
  pages={1482-1495},
  doi={10.1109/HPCA61900.2025.00111}}

@INPROCEEDINGS{A3,
  author={Ham, Tae Jun and Jung, Sung Jun and Kim, Seonghak and Oh, Young H. and Park, Yeonhong and Song, Yoonho and Park, Jung-Hun and Lee, Sanghee and Park, Kyoung and Lee, Jae W. and Jeong, Deog-Kyoon},
  booktitle={2020 IEEE International Symposium on High Performance Computer Architecture (HPCA)}, 
  title={A3: Accelerating Attention Mechanisms in Neural Networks with Approximation}, 
  year={2020},
  volume={},
  number={},
  pages={328-341},
  doi={10.1109/HPCA47549.2020.00035}}

@inproceedings{MX_PLUS,
author = {Lee, Jungi and Park, Junyong and Cha, Soohyun and Cho, Jaehoon and Sim, Jaewoong},
title = {MX+: Pushing the Limits of Microscaling Formats for Efficient Large Language Model Serving},
year = {2025},
isbn = {9798400715730},
publisher = {Association for Computing Machinery},
address = {New York, NY, USA},
url = {https://doi.org/10.1145/3725843.3756118},
doi = {10.1145/3725843.3756118},
booktitle = {Proceedings of the 58th IEEE/ACM International Symposium on Microarchitecture},
pages = {869-883},
numpages = {15},
location = {
},
series = {MICRO '25}
}

@inproceedings{amxfp4,
    title = "{AMXFP}4: Taming Activation Outliers with Asymmetric Microscaling Floating-Point for 4-bit {LLM} Inference",
    author = "Lee, Janghwan  and
      Park, Jiwoong  and
      Kim, Jinseok  and
      Kim, Yongjik  and
      Oh, Jungju  and
      Oh, Jinwook  and
      Choi, Jungwook",
    editor = "Che, Wanxiang  and
      Nabende, Joyce  and
      Shutova, Ekaterina  and
      Pilehvar, Mohammad Taher",
    booktitle = "Findings of the Association for Computational Linguistics: ACL 2025",
    month = jul,
    year = "2025",
    address = "Vienna, Austria",
    publisher = "Association for Computational Linguistics",
    url = "https://aclanthology.org/2025.findings-acl.776/",
    doi = "10.18653/v1/2025.findings-acl.776",
    pages = "14993--15013",
    ISBN = "979-8-89176-256-5"
}

@misc{micromix,
      title={MicroMix: Efficient Mixed-Precision Quantization with Microscaling Formats for Large Language Models}, 
      author={Wenyuan Liu and Haoqian Meng and Yilun Luo and Yafei Zhao and Peng Zhang and Xindian Ma},
      year={2026},
      eprint={2508.02343},
      archivePrefix={arXiv},
      primaryClass={cs.LG},
      url={https://arxiv.org/abs/2508.02343}, 
}

@ARTICLE{ramulator,
  author={Luo, Haocong and Tuğrul, Yahya Can and Bostancı, F. Nisa and Olgun, Ataberk and Yağlıkçı, A. Giray and Mutlu, Onur},
  journal={IEEE Computer Architecture Letters}, 
  title={Ramulator 2.0: A Modern, Modular, and Extensible DRAM Simulator}, 
  year={2024},
  volume={23},
  number={1},
  pages={112-116},
  doi={10.1109/LCA.2023.3333759}}
\end{document}